\documentclass[preprint]{revtex4}

\usepackage[dvipdfmx]{graphicx}
\usepackage{feynmf}
\usepackage{amsmath}
\usepackage{amssymb}
\usepackage{here}
\usepackage{mathrsfs}
\usepackage{color}
\usepackage{float}

\def\vec#1{\mbox{\boldmath $#1$}}
\def\Slash#1{\setbox0=\hbox{$#1$}               
   \dimen0=\wd0                                 
   \setbox1=\hbox{/} \dimen1=\wd1               
   \ifdim\dimen0>\dimen1                        
      \rlap{\hbox to \dimen0{\hfil/\hfil}}      
      #1                                        
   \else                                        
      \rlap{\hbox to \dimen1{\hfil$#1$\hfil}}   
      /                                         
   \fi}

\def\D{{\mathcal D}}

\def\J{{\mathcal J}}

\def\S{{\mathcal S}}
\def\V{{\mathcal V}}

\begin{document}

\title{
 Sensitivity of indirect detection of Neutralino dark matter by Sommerfeld enhancement mechanism 
}
\author{Mikuru Nagayama}
\email{nagayama@krishna.th.phy.saitama-u.ac.jp}
\affiliation{%
Department of Physics, Saitama University,
\\
Shimo-Okubo 255, 
338-8570 Saitama Sakura-ku, 
Japan
}

\author{Joe Sato}
\email{joe@phy.saitama-u.ac.jp}
\affiliation{%
Department of Physics, Saitama University,
\\
Shimo-Okubo 255, 
338-8570 Saitama Sakura-ku, 
Japan
}

\author{Yasutaka Takanishi}
\email{yasutaka@krishna.th.phy.saitama-u.ac.jp}
\affiliation{%
Department of Physics, Saitama University,
\\
Shimo-Okubo 255, 
338-8570 Saitama Sakura-ku, 
Japan
}

\author{Kazuhiro Tsunemi}
\affiliation{%
Department of Physics, Saitama University,
\\
Shimo-Okubo 255, 
338-8570 Saitama Sakura-ku, 
Japan
}

\

\begin{abstract}
  We have investigated neutralino dark matter in the framework of
  minimal supersymmetric Standard Model focusing on the
  coannihilatioin region. In this region, where the particle whose
  mass is tightly degenerated with the neutralino dark matter exists,
  we can solve the Lithium problem in the case of lepton flavor being
  violated.  It turns out that Sommerfeld enhancement is important in
  the coannihilation region so that the dark matter signal becomes
  large enough to be observed by the current sensitivity of indirect
  experiments.

\end{abstract}

\date{\today}



\preprint{\bf }

\maketitle

\section{Introduction}
The discovery of neutrino oscillation in 1998 opens a new era for
physics Beyond the Standard Model (SM)~\cite{Fukuda_1998}.
After this discovery, the flavor structure of the lepton sector
must be considered and its research becomes a very important topic.
On the other hand, another road of Beyond the Standard Model
is a quest for Dark Matter (DM) problem. 
The existence of DM is according
to various astrophysical observations, including gravitational effects
on the visible matter in the infrared and gravitational lensing of
background radiation~\cite{Wittman:2000tc}. Furthermore, the predictions of the SM of
cosmology are confirmed by various cosmological observations namely
the total mass-energy of
Universe contains about 23$\%$ DM and
68$\%$ of a form of dark energy so that our present Universe contains
only about 4$\%$ of the ``ordinary'' matter and energy. However, we do
not really know what DM is.

The total abundance of DM, which has important implications for the
evolution of the Universe, has been precisely measured by the WMAP
collaboration~\cite{Komatsu:2010fb} during the last few decades. This
requires that a different kind of matter beyond the SM of particle
physics must be considered. One of the most popular and most intensive
studied candidates is the so-called weakly interacting massive particle
(WIMP) that may constitute most of the matter in the
Universe~\cite{Aghanim:2018eyx}. Cosmology provides, therefore, a good
motivation for Supersymmetry (SUSY) that has a natural candidate for
DM \mbox{\it i.e.}  the lightest supersymmetric particle (LSP) can be
a stable particle if R-parity conservation is held~\cite{Ellis:1983ew}
(for reviews see \mbox{\it e.g.}~\cite{Jungman_1996}). As many other
physicists have already suggested and studied the scenarios of
lightest neutralino being the candidate for LSP in various different
frameworks of SUSY models.

We assume that the dark matter particle is neutralino, and we focus on
the so-called coannihilation region. In such a region, the mass
between LSP neutralino and next-LSP(NLSP) slepton degenerates, we are
able to reduce the abundance of DM by an order or more orders of
magnitude with a fixed value of DM mass. In addition, Lithium
problem~\cite{Ryan:2000zz,Sbordone:2010zi,Cyburt:2008kw} can be solved
if we admit lepton flavor
violation~\cite{Jittoh:2007fr,Jittoh:2008eq,Jittoh:2010wh,Jittoh:2011ni,Kohri:2012gc,Konishi:2013gda}. For
this reason Sommerfeld
Enhancement~\cite{Hisano_2007,Ellis:1999mm,ArkaniHamed:2008qn,Feng:2010zp}
effect plays a crucial role in indirect DM observation experiments
such as HESS experiment and Fermi-LAT
experiment~\cite{Abramowski:2014tra, Abdo:2009zk}. In fact, this
effect has been used in many studies that referred to indirect
detection~\cite{Biondini:2018ovz,ElAisati:2017ppn,Abdalla:2018mve,Braaten:2017kci}.
The phenomenon is that the annihilation cross section is enhanced by
forming a bound state when DMs move non-relativistically and
annihilate into SM particles~\cite{Lisanti:2016jxe}. Thus our results
bring attention to current and future experiments of astrophysics.

This article is organized as follows: in the next section, we define
the Lagrangian related to neutralino and the lightest slepton, and we
clarify our notation. In section 3, the method of two-body effective
action is reviewed and discussed. Calculations of cross sections and
fluxes from dark matter annihilation are also given. In section 4, our
calculation results will be presented and discussed. Finally, we
conclude in section 5. We summarize the technical details in Appendices A, B, and C.

\section{Lagrangian}
\unitlength = 1mm \indent\ %
In this section, we consider the neutralino-slepton coannihilation
region in the framework of  Minimal Supersymmetric Standard Model (MSSM). Here we assume that DM is LSP Bino-like
neutralino and NLSP are the lightest slepton which is required
very tight mass degeneracy with LSP neutralino.  It is important to
note that we use  terminology in this article that the lightest
slepton is ``stau'' because $\widetilde\ell_1$ almost consists
of right-handed stau in the flavor base.

It is convenient to use a mass base for calculation, so we begin by
formulating Lagrangian with this base.  In MSSM, neutralino is a linear
combination of Higgsino neutral component of ($\widetilde{H}_u^0,\widetilde{H}_d^0$), Wino
neutral component $(\widetilde{W}^0)$, and Bino
$(\widetilde{B}^0)$. In this base neutralino mass matrix is given by
\begin{align}
\mathscr{L}_{\rm neutralino \ mass}=-\frac{1}{2} (\widetilde{\psi}^0)^{\sf T} M_N \widetilde{\psi}^0 + c.c.
\end{align}
Here  $\widetilde{\psi}^0=(\widetilde{B},\widetilde{W}^0,\widetilde{H}_u^0,\widetilde{H}_d^0)^{\sf T}$, and 
\begin{eqnarray}
M_N=\left(
\begin{array}{ccccccc}
M_1  && 0    && -c_\beta s_w m_z &&  s_\beta s_w m_z \\
0    && M_2  &&  c_\beta c_w m_z && -s_\beta c_w m_z \\
-c_\beta s_w m_z &&  c_\beta c_w m_z && 0 && -\mu     \\
 s_\beta s_w m_z && -s_\beta c_w m_z && -\mu && 0   \\  
\end{array}
\right) \hspace{2mm},
\end{eqnarray}
where $M_1$ is Bino mass, $M_2$ is Wino mass, $\mu$ is Higgsino mass,
$m_z$ is $Z$ boson mass, respectively. $\theta_w$ is Weinberg angle,
and we use shorthand notation: $s_w=\sin{\theta_w}$ and
$c_w=\cos{\theta_w}$. We also denote $\tan{\beta}=v_u/v_d$, where
$v_u$ and $v_d $ are vacuum expectation values of $H_u$ and $H_d$,
respectively. Also $s_\beta =\sin{\beta}$ and $c_\beta=\cos{\beta}$.

The $\widetilde{\psi}^0$ base can be transformed into mass base
$\widetilde{\chi}$ using unitary matrix
$\displaystyle \left(N_{\widetilde{G}}\right)_a{}^b~(a,b=1, \cdots,
4)$,
\begin{align}
{\widetilde{\chi}}{_a} = \left(N_{\widetilde{G}}\right)_a{}^b~ \widetilde{\psi}^0{_b}.
\end{align}
In this article we do consider only the case of the lightest
neutralino being the candidate of DM, thus we fix the value of $a=1$.

The existence of mass degenerated particle with DM in the coannihilation
process is necessary, thus we assume such particle being the lightest
slepton. This slepton mass matrix follows
\begin{align}
\mathscr{L}_{\rm slepton \ mass}=-\widetilde{\psi}_l^\dagger ~M_{\widetilde{l}}^2 ~\widetilde{\psi}_l 
\end{align}
in flavor base $\widetilde{\psi}_l=(\widetilde{e}{_L} \ \ \widetilde{\mu}{_L} \ \ \widetilde{\tau}{_L} \
\ \widetilde{e}_R \ \ \widetilde{\mu}_R \ \ \widetilde{\tau}_R)^{\sf T}$, and $M_{\widetilde{l}}^2$ is given by
\begin{equation}
{(M_{\widetilde{l}}^2)_I}^J=
   \left\{ \begin{array}{l r}
 (m_L^2){_I}{^J} + y{^\dagger}{_I}^K y{_K}^J v_d^2 + m_z^2 (s_w^2-1/2)c_{\beta} \delta{_I}{^J} & (\text{for} \,  I,J=1,2,3, K=4,5,6)\\
  -\mu v_u y{^\dagger}{_I}^J +v_d a{^\dagger}{_I}^J   \qquad \qquad \qquad & (\text{for} \, I= 1,2,3, J=4,5,6) \\
 -\mu^* v_u y{_I}^J +v_d a{_I}^J   \qquad \qquad \qquad & (\text{for} \, I=4,5,6, J= 1,2,3) \\
  (m_R^2){_I}{^J} + y{_I}^K y{^\dagger}{_K}^J v_d^2 + m_z^2 s_w^2 c_{\beta} \delta{_I}{^J} \hspace{2mm}  & (\text{for} \, I,J= 4,5,6, K=1,2,3),
  \end{array}\right. 
\end{equation}
where $m_L^2,m_R^2$ is soft-breaking mass parameter. ${y_I}^J$ is
Yukawa coupling and ${a_I}^J$ is occurred from A-term, where
$I (I=4,5,6)$ represents right-hand subscript and $J (J=1,2,3)$
represents left-hand superscript, respectively. Note that if we take
Hermitian conjugate, right-hand and left-hand are reversed. This base
is also changed mass base $\widetilde{l}$ using unitary matrix
$N_{\widetilde{l}}{_A}^B\ (A,B=1,\cdots, 6)$,
\begin{equation}
\widetilde{l}_A=N_{\widetilde{l}}{_A}^B \widetilde{\psi}_l{_B} \,,
\end{equation} 
Especially, we assume that the lightest slepton $\widetilde{l}_1$ almost consists of stau $\widetilde{\tau}$, so
we note $\widetilde{l}_1$ as $\widetilde{\tau}$.

In the same way, we can write down lagrangian of Chargino in the mass
base. Chargino is linear combination of Higgsino charged components
$(\widetilde{H}_d^-$, $\widetilde{H}_u^+)$ and Wino charged components
$\displaystyle \left(\widetilde{W}^+,\widetilde{W}^-\right)$.  The
mass matrix is written
\begin{equation}
\mathscr{L}_{\rm chargino \, mass}=-\frac{1}{2} (\widetilde{\psi}^\pm)^{\sf T}M_C \widetilde{\psi}^\pm + h.c.
\end{equation}
in $\widetilde{\psi}^\pm=(\widetilde{W}^+,\widetilde{H}_u^+,\widetilde{W}^-,\widetilde{H}_d^-)^{\sf T}$ base. Here $M_C$ is
\begin{eqnarray*}
M_C=\left(
\begin{array}{ccc}
0  && X^{\sf T} \\
X && 0 \\
\end{array}
\right)\hspace{2mm}, \hspace{1cm} \text{where} \hspace{3mm}
X=\left(
\begin{array}{ccc}
M_2  && \sqrt{2} s_\beta m_w \\
\sqrt{2} c_\beta m_w && \mu \\
\end{array}
\right)\, .
\end{eqnarray*}
By using two unitary matrices $U$ and $V$, the mass base becomes
\begin{eqnarray}
\left(
\begin{array}{c}
\widetilde{C}_1^+ \\
\widetilde{C}_2^+ \\
\end{array}
\right)
=V\left(
\begin{array}{c}
\widetilde{W}^+ \\
\widetilde{H}_u^+ \\
\end{array}
\right), \qquad \left(
\begin{array}{c}
\widetilde{C}_1^- \\
\widetilde{C}_2^- \\
\end{array}
\right)
=U\left(
\begin{array}{c}
\widetilde{W}^-\\
\widetilde{H}_d^- \\
\end{array}
\right).
\end{eqnarray}

The main contributions of Lagrangian which appear in the below diagram are
described in the following in terms of the above fields (for reader
interest in all interaction, see Appendix A.).
\begin{align}
\label{lagrangian}
\mathscr{L} &= \mathscr{L}_{\rm KT} + \mathscr{L}_{\rm int}  \nonumber \\
   & \sim \frac{1}{2}\bar{\widetilde{\chi}}\left(i\Slash{\partial}-m\right)\widetilde{\chi} +{\bar{e}}^i \left(i\Slash{\partial}{\delta_i}^j-{(m_e)_i}^j\right)e_j -\widetilde{\tau}^*(\partial^2+m_{\widetilde{\tau}}^2)\widetilde{\tau} \nonumber \\ 
&+\frac{1}{2}Z_\mu (\partial^2+m_Z^2)Z^\mu+\frac{1}{2}A_\mu \partial^2A^\mu + \mathscr{L}_{\rm gauge} + \cdots \,,
\end{align}
where $i,j = 1,2,3$ and
$\displaystyle \left({m_e}\right)_i{^j} = {\rm diag}(m_e , m_\mu, m_\tau)$. We pick up $\mathscr{L}_{\rm gauge}$ as an example of $\mathscr{L}_{\rm int}$ which is shown in Eq.~(\ref{Lag}). $\mathscr{L}_{\rm gauge}$ is written in the following form.
\begin{align}
\mathscr{L}_{\rm gauge} = & i e A_\mu \widetilde{\tau}^* \overleftrightarrow{\partial}^\mu  \widetilde{\tau} 
-i g_z Z_\mu \left(s_w^2-\frac{1}{2} N_{\widetilde{l}}{_1}^i N^\dagger_{\widetilde{l}}{_i}^1\right) \widetilde{\tau}^* \overleftrightarrow{\partial}^\mu  \widetilde{\tau} \nonumber \\
&-i\frac{\sqrt{2}}{2}g \left( W_\mu ^+ \widetilde{\nu}{^*}{^i} \overleftrightarrow{\partial}^\mu N^\dagger_{\widetilde{l}}{_i}^1 \widetilde{\tau}
+W_\mu ^- \widetilde{\tau}^* N_{\widetilde{l}}{_1}^i \overleftrightarrow{\partial}^\mu  \widetilde{\nu}{_i} \right) \nonumber \\
&+e^2 A^2 |\widetilde{\tau}|^2 + g_z^2 \left(s_w^2-\frac{1}{2} N_{\widetilde{l}}{_1}^i N^\dagger_{\widetilde{l}}{_i}^1\right)^2 Z^2 |\widetilde{\tau}|^2 -2eg_z \left(s_w^2-\frac{1}{2} N_{\widetilde{l}}{_1}^i N^\dagger_{\widetilde{l}}{_i}^1\right) A_\mu Z^\mu |\widetilde{\tau}|^2 \nonumber \\
&+\frac{g^2}{2} N_{\widetilde{l}}{_1}^i N^\dagger_{\widetilde{l}}{_i}^1 W_\mu ^+ W^{- \mu} |\widetilde{\tau}|^2 \,,
\end{align}
where the sum of $i$ is taken from 1 to 3, and  we have used the following definition 
\begin{align}
\widetilde{\tau}^* \overleftrightarrow{\partial}^\mu  \widetilde{\tau} = \widetilde{\tau}^* \partial^\mu \widetilde{\tau} - (\partial^\mu \widetilde{\tau}^*) \widetilde{\tau} \hspace{2mm}.
 \end{align}
Here we note $g$ and $g'$ as $SU(2)$ and $U(1)$ coupling respectively, and $g = c_w g_z$.

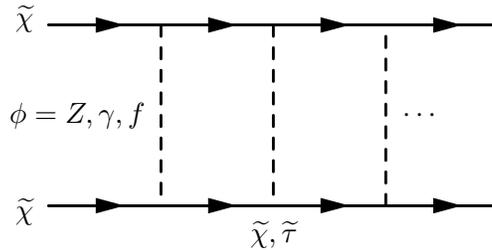
\begin{figure}[H]
\centering
\begin{fmffile}{box}
\parbox{60mm}{\begin{fmfgraph*}(60,30)
\fmftop{i1,o1}
\fmfbottom{i2,o2}
\fmflabel{$\widetilde{\chi}$}{i1}
\fmflabel{$\widetilde{\chi}$}{i2}
\fmf{fermion,tension=0.5}{i1,v1}
\fmf{fermion,tension=0.5}{v1,v2}
\fmf{fermion,tension=0.5}{v2,v3}
\fmf{fermion,tension=0.5}{v3,o1}
\fmf{fermion,tension=0.5}{i2,v4}
\fmf{fermion,tension=0.5}{v4,v5}
\fmf{fermion,tension=0.5}{v5,v6}
\fmf{fermion,tension=0.5}{v6,o2}
\fmf{dashes,label=$\phi=Z,,\gamma,,f$,tension=0}{v1,v4}
\fmf{dashes,tension=0}{v2,v5}
\fmf{dashes,tension=0,label=$\cdots$,l.side=right}{v6,v3}
\fmflabel{$\widetilde{\chi},\widetilde{\tau}$}{v5}
\end{fmfgraph*}}
\end{fmffile}
\caption{DM annihilation ladder diagram. $Z,\gamma$ and $f$ are taken as an example of $\phi$.}
\end{figure}

\section{Formalism}
\subsection{Two-body effective action}
\indent\ 
In this section, we derive non-relativistic two-body effective
action that has been investigated in Ref.~\cite{Hisano:2004ds}. 
We will apply their method to calculate the
cross sections as of our interest. The steps are as follows:
\mbox{({i})} We integrate out the fields except for
$\widetilde{\chi},\widetilde{\tau}$. \mbox{({ii})} We integrate out
large momentum mode of $\widetilde{\chi}$ and
$\widetilde{\tau}$. \mbox{({iii})} The non-relativistic action obtained
in~\mbox{({ii})} is expanded by DM velocity. \mbox{({iv})} We
introduce auxiliary fields that represent a two-body state and
integrate out all fields except these auxiliary fields.

We begin by integrating out the fields except for
$\widetilde{\chi},\widetilde{\tau}$ and obtain 1-loop effective
action. For example, we consider integrating out $A_\mu$. First, we
choose the terms related to $A_\mu$ from Eq.~(\ref{lagrangian}), and
get
\begin{align}
\label{Amu}
  \S_A= & -i \ln\int DA \exp i \left[\int d^4 x\left(\frac{1}{2}A_\mu \partial^2g^{\mu \nu}A_\nu
          + i e A_\mu \widetilde{\tau}^* \overleftrightarrow{\partial}^\mu  \widetilde{\tau} \right. \right. \nonumber\\
       &\qquad \qquad \qquad \qquad \qquad \left. \left. + e^2 A^2 \left|\widetilde{\tau}\right|^2
         -2eg_z \left(s_w^2-\frac{1}{2} N_{\widetilde{l}}{_1}^i N^\dagger_{\widetilde{l}}{_i}^1\right) A_\mu
         Z^\mu \left|\widetilde{\tau} \right|^2\right)\right],
\end{align}
where the sum of $i$ is taken from 1 to 3. Next, we replace $A_\mu$ by
\begin{gather*}
A_\mu \rightarrow A_\mu + i\int d^4y {{\cal D}_{\mu \nu}^A}(x-y) {\cal J}^\nu(y) \hspace{2mm}
\end{gather*}
and use the Green function's relation
\begin{align}
\mathscr{L}^{\mu \nu}_A {\D_{\nu \rho}^A}(x-y) = i\delta(x-y) \delta^\mu _\rho.
\end{align}
Then Eq.~(\ref{Amu}) becomes
\begin{align}
\label{Amu2}
\S_A= -i \ln {\rm Det}^{-\frac{1}{2}} (-\mathscr{L}^{\mu \nu}_A )
+ \frac{i}{2}\int d^4 x d^4 y \J^\mu(x) {\D_{\mu \nu}^A}(x-y) \J^\nu(y) \hspace{2mm},
\end{align}
where
\begin{align*}
\mathscr{L}^{\mu \nu}_A =& (\partial^2 +2e^2 |\widetilde{\tau}|^2)g^{\mu \nu}, \nonumber \\
\J^\mu(x) = &  i e  \widetilde{\tau}^* \overleftrightarrow{\partial}^\mu  \widetilde{\tau}  -2eg_z  \left(s_w^2-\frac{1}{2}  {N_{\widetilde{l} 1}}^i {N^\dagger_{\widetilde{l} i}}^1\right)Z^\mu |\widetilde{\tau}|^2. \nonumber \\
\end{align*}
Here, we replace the first term of Eq.~(\ref{Amu2}) as follows.
\begin{align}
-i \ln {\rm Det}^{-\frac{1}{2}} (-\mathscr{L}^{\mu \nu}_A )
= & \frac{i}{2}{\rm Tr} \left[ \ln \left( -\partial^2 - 2e^2 |\widetilde{\tau}|^2)g^{\mu \nu} \right) \right] \nonumber \\
\equiv & ~\frac{i}{2}{\rm Tr} \left[ \ln \left( A_0 + \delta A \right) \right] \nonumber \\
= & ~\frac{i}{2}{\rm Tr} \left[ \ln A_0 + A_0^{-1} \delta A -\frac{1}{2} A_0^{-1} \delta A A_0^{-1} \delta A \right] \nonumber \\
\sim & i e^4 {\rm tr} \int d^4 x_1 d^4 x_2 |\widetilde{\tau}|^2(x_1) |\widetilde{\tau}|^2(x_2) {D_{\mu \nu}^A}(x_1-x_2)  D^{A \nu \rho }(x_2-x_1) \,.
\end{align}
Note that Tr() donates operator trace and tr() donates Dirac trace. Also $D_{\mu \nu}^A(x-y)$ represents photon propagator,
\begin{align}
A_0^{-1} = i {D_{\mu \nu}^A}(x-y) = i \int \frac{d^4q}{{(2\pi)}^4}\frac{-i g_{\mu \nu}}{q^2+i\epsilon}e^{-iq(x-y)}.
\end{align}
The second term of Eq.~(\ref{Amu2}), $\D_{\mu \nu}^A \sim D_{\mu \nu}^A$ at the lowest order of expansion. Finally, we get effective action on $A_\mu$ as follows
\begin{align}
\S_A=& i e^4 {\rm tr} \int d^4 x_1 d^4 x_2 |\widetilde{\tau}|^2(x_1) |\widetilde{\tau}|^2(x_2) {D_{\mu \nu}^A}(x_1-x_2)D^{A \nu \rho }(x_2-x_1) \nonumber \\
&+ \frac{i}{2}\int d^4 x d^4 y {\cal J}^\mu(x) {D_{\mu \nu}^A}(x-y) {\cal J}^\nu(y) .
\end{align}
By this calculation, we can represent the 1-loop interaction shown in Fig.~2.
 
\begin{figure} [t]
\label{photon-1loop}
\centering
\begin{fmffile}{optgamma}
\parbox{30mm}{\begin{fmfgraph*}(40,30)
\fmfleft{i1,i2}
\fmfright{o1,o2}
\fmf{dashes_arrow,label=$\widetilde{\tau}^*$}{v1,i1}
\fmf{dashes_arrow,label=$\widetilde{\tau}$}{i2,v1}
\fmf{photon,label=$\gamma$,left=0.7,tension=0.5}{v1,v2}
\fmf{photon,label=$\gamma$,right=0.7,tension=0.5}{v1,v2}
\fmf{dashes_arrow,label=$\widetilde{\tau}^*$}{o1,v2}
\fmf{dashes_arrow,label=$\widetilde{\tau}$}{v2,o2}
\end{fmfgraph*}}
\end{fmffile}
\caption{1-loop interaction diagram mediated by photons.}
\end{figure}
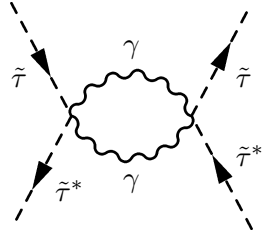
 
After all fields except $\widetilde{\chi},\widetilde{\tau}$ are integrated out, the effective action becomes 
\begin{align}
\S_{eff}=&\int d^4x\left[ \frac{1}{2}\bar{\widetilde{\chi}}\left(i\Slash{\partial}-m\right)\widetilde{\chi}
-\widetilde{\tau}^*(\partial^2+m_{\widetilde{\tau}}^2)\widetilde{\tau} \right] \nonumber \\
&+{\cal S}_A'+{\cal S}_Z+{\cal S}_W'+{\cal S}_{\widetilde{\nu}}+{\cal S}_e+{\cal S}_{\widetilde{C}}+{\cal S}_{h^0}.
\end{align}
One can see other action of $\S_A$ in Appendix B.

Next, we integrate out the large momentum modes of $\widetilde{\chi},\widetilde{\tau}$. Here we divide the fields two parts, namely relativistic part and non-relativistic part. For the case of $\widetilde{\chi}$, it will be
\begin{align}
\widetilde{\chi}(x) =& \widetilde{\chi}(x)_R + \widetilde{\chi}(x)_{NR}, \nonumber \\
\widetilde{\chi}(x)_R =& \int_R \frac{d^4q}{(2\pi)^4} \phi^0(q) e^{-iqx}, \nonumber \\
\widetilde{\chi}(x)_{NR} =& \int_{NR} \frac{d^4q}{(2\pi)^4} \phi^0(q) e^{-iqx},
\end{align}
where $\phi^0$ is the Fourier coefficient of the DM field. After this division, we integrate out $\widetilde{\chi}_R$. The same operation is done for $\widetilde{\tau}$, in the result we obtain
\begin{align}
\S_{NR}=\int d^4x \left[ \frac{1}{2} \bar{\widetilde{\chi}}_{NR}(i\Slash{\partial}-m)\widetilde{\chi}_{NR} - \widetilde{\tau}_{NR}^* (\partial^2 + m_{\widetilde{\tau}}^2)\widetilde{\tau}_{NR} \right]+\S_{Pot}(\widetilde{\chi}_{NR},\widetilde{\tau}_{NR})+\S_{Im}(\widetilde{\chi}_{NR},\widetilde{\tau}_{NR}),
\end{align}
Here we note $\S_{Pot}$ is the real part except the kinematic part, and $\S_{Im}$ is the imaginary part. In the following, we omit the subscript NR for $\widetilde{\chi}$ and $\widetilde{\tau}$.

Then, we expand this action by DM velocity. For this expansion, we use two-component spinors of neutralino and stau. These spinor fields are defined in the following form.
\begin{align}
\bar{\chi} = \left( \begin{array}{c} e^{-imt}\zeta + ie^{imt} \frac{\overrightarrow{\nabla} \cdot \sigma}{2m} \zeta^c\\ 
e^{imt}\zeta^c - ie^{-imt} \frac{\overrightarrow{\nabla} \cdot \sigma}{2m} \zeta \end{array} \right), \qquad
\widetilde{\tau} = \frac{1}{\sqrt{2m}}\eta e^{-imt} + \frac{1}{\sqrt{2m}} \xi e^{imt},
\end{align}
where $\zeta^c = -i\sigma^2{\zeta^\dagger}^{\sf T}$. In this form, $\S_{NR}$ becomes
\begin{align}
\S_{NR} = \S_{KT} + \S_{Pot} + \S_{Im} ,
\end{align}
where 
\begin{align}
\S_{KT}=&\frac{1}{2} \int d^4 x \bar{\widetilde{\chi}}(x)(i\Slash{\partial}-m) \widetilde{\chi}(x) - \int d^4 x \widetilde{\tau}^* (\partial^2 + m_{\widetilde{\tau}}^2)\widetilde{\tau} \nonumber \\
=& \int d^4 x \left[\zeta^\dagger \left( i \partial_0 + \frac{\nabla^2}{2m}\right)\zeta + \eta^*\left( i\partial_0 + \frac{\nabla^2}{2m} - \delta m \right)\eta - \xi^*\left( i\partial_0 - \frac{\nabla^2}{2m} + \delta m \right)\xi \right].
\end{align}
We define to quantify $\delta m$ as
\begin{equation}
\delta m = \frac{{m_{\widetilde{\tau}}}^2-m^2}{2m},
\end{equation}
and the potential part, 
\begin{align}
\S_{Pot}=&\int d^4 x d^4 y  \delta(x^0-y^0) \left[\frac{\alpha }{r}
+g_z^2\left(s_w^2-\frac{1}{2} N_{\widetilde{l}}{_1}^i N^\dagger_{\widetilde{l}}{_i}^1\right)^2 \frac{e^{-m_z r}}{4 \pi r}
+C_{h^0}^2 \frac{e^{-(m_{h^0}) r}}{8 \pi m^2} \right]
\eta^*(x) \eta(x) \xi^*(y) \xi(y) \nonumber \\
&+\int d^4 x d^4 y \frac{e^{-m_{e_i} r} \delta(x^0-y^0)}{16 \pi m r} \left(C^i_1 {{(m_e)}_i}^j {C^\dagger}_{2j} - C^i_2 {{(m_e)}_i}^j {C^\dagger}_{1j}\right) \nonumber \\
&\times \left(\eta ^* (x) {\zeta^c}^\dagger (x) \xi (y) \zeta (y) - \xi^* (x) \zeta^\dagger (x) \eta (y) \zeta^c (y)\right)\nonumber \\
\equiv& \int d^4 x d^4 y  \delta(x^0-y^0) \left[ {\cal S}_{Pot}^{(1)} \eta^*(x) \eta(x) \xi^*(y) \xi(y) \right.\nonumber \\
&+\left. {\cal S}_{Pot}^{(2)} \left(\eta ^* (x) {\zeta^c}^\dagger (x) \xi (y) \zeta (y) - \xi^* (x) \zeta^\dagger (x) \eta (y) \zeta^c (y) \right) \right],
\end{align}
where $i,j=1,2,3$ and the sum of $i$ is taken from 1 to 3. Note that $C_1^i , C_2^i$ and $C_{h^0}$ are defined as in Eq.~(\ref{C1}) , Eq.~(\ref{C2}), and Eq.~(\ref{Ch0}), respectively, in Appendix B. Also, the imaginary part becomes 
\begin{align}
\S_{Im} = \S_{\gamma} +  \S_{e} + \S_{Z}+  \S_{Z h^0} +  \S_{h^0} +  \S_{W} +  \S_{\nu} .
\end{align}
For example, $\S_{\gamma}$ is calculated the diagram drawn above (see Fig.~2) by optical theorem, then we obtain 
\begin{align}
\S_{\gamma}  =  &  i \frac{e^4}{8 \pi m^2} \ \int d^4 x \eta^*(x) \eta(x)\xi^*(x)\xi(x) \nonumber \\
\equiv & i \Gamma_{\gamma \gamma} \int d^4 x \eta^*(x) \eta(x)\xi^*(x)\xi(x).
\end{align}
One can see the other actions except for $\S_\gamma$ in Appendix C.

Next, we introduce auxiliary fields $\sigma_{\widetilde{\chi}},\sigma_{\widetilde{\tau}}$ to make two-body states  $\widetilde{\chi} \widetilde{\chi}$ and $\widetilde{\tau} \widetilde{\tau}$, which satisfy the relationship:
\begin{align*}
1=&\int D\sigma_{\widetilde{\tau}} Ds_{\widetilde{\tau}}^\dagger \exp \left[ \frac{i}{2} \int d(xy) \sigma_{\widetilde{\tau}} (t,\vec{x},\vec{y}) \left( s_{\widetilde{\tau}}^\dagger (t,\vec{x},\vec{y}) -i \eta^*(t,\vec{x}) \xi(t,\vec{y}) \right) \right], \\
1=&\int D\sigma_{\widetilde{\tau}}^\dagger Ds_{\widetilde{\tau}} \exp \left[ \frac{i}{2} \int d(xy) \sigma_{\widetilde{\tau}}^\dagger  (t,\vec{x},\vec{y}) \left( s_{\widetilde{\tau}} (t,\vec{x},\vec{y}) -i \xi^*(t,\vec{y}) \eta(t,\vec{x}) \right) \right], \\
1=&\int D\sigma_{\widetilde{\chi}} Ds_{\widetilde{\chi}}^\dagger \exp \left[ \frac{i}{2} \int d(xy) \sigma_{\widetilde{\chi}} (t,\vec{x},\vec{y}) \left( s_{\widetilde{\chi}}^\dagger (t,\vec{x},\vec{y}) -\frac{1}{2} \zeta^\dagger(t,\vec{x}) \zeta^c(t,\vec{y}) \right) \right], \\
1=&\int D\sigma_{\widetilde{\chi}}^\dagger Ds_{\widetilde{\chi}} \exp \left[ \frac{i}{2} \int d(xy) \sigma_{\widetilde{\chi}}^\dagger (t,\vec{x},\vec{y}) \left( s_{\widetilde{\chi}} (t,\vec{x},\vec{y}) -\frac{1}{2} {\zeta^c}^\dagger(t,\vec{y}) \zeta(t,\vec{x}) \right) \right] ,
\end{align*}
where $d(xy)=dtd^3xd^3y$. With this relationship, we integrate out the fields except for $\sigma_{\widetilde{\chi}}$ and $\sigma_{\widetilde{\tau}}$. Then two-body state effective action is obtained as
\begin{align}
\label{nonRela}
\S^{I\hspace{-.1em}I}=\int d^4x d^3r  \Phi^\dagger (x,\vec{r}) \left( \frac{\nabla_r^2}{m} + i \partial_{x^0} + \frac{\nabla_x^2}{4m}
-\left(
\begin{array}{ccc}
2\delta m && 0\\
0 && 0\\
\end{array}
\right)
+\frac{1}{2}\left(
\begin{array}{ccc}
-1 && 0\\
0 && 1\\
\end{array}
\right)
{\cal V} \right) \Phi(x,\vec{r}),
\end{align}
where $\Phi(x,\vec{r})$ is represented as
\begin{align}
\Phi (x, \vec{r}) = \V^{-1} \left( \begin{array}{l} \sigma_{\widetilde{\tau}}(x,\vec{r}) \\ \sigma_{\widetilde{\chi}}(x,\vec{r})\end{array} \right),  \quad \V=\left( \begin{array}{cc}  
-2\S_{Pot}^{(1)} - 2i\delta(\vec{x}-\vec{y})\Gamma & ~~ -4i\S_{Pot}^{(2)} \\
 4i\S^{(2)}_{Pot} & 0 
\end{array}\right).
\end{align}
Note that $x$ denotes the center of mass coordinate in two-body system and $\vec{r}$ is the relative  coordinate, and
here we note $\Gamma$ as 
\begin{align}
\label{Gamma}
\Gamma=\Gamma_{e_ie_i}+\Gamma_{\gamma \gamma}+\Gamma_{Z^0Z^0} +\Gamma_{Z^0\gamma}+\Gamma_{Z^0 h^0}+\Gamma_{h^0 h^0}+\Gamma_{\nu_i \nu_i}.
\end{align}
The symbols $\Gamma_f$ ($f=e_ie_i,\gamma \gamma,Z^0Z^0,Z^0\gamma,Z^0h^0,h^0 h^0,\nu_i \nu_i$) are listed in Appendix C. 

\subsection{cross section}
In this subsection, we calculate  DM annihilation cross section with the method written in \cite{Hisano:2004ds,Matsumoto:2005ui}.
For S-wave, annihilation cross section is obtained by determining radial component of Green function which satisfies Schwinger-Dyson equation derived from Eq.~(\ref{nonRela}):
\begin{align}
\label{schwinger} 
\left[ \frac{\nabla_r^2}{m} + i \partial_{x^0} + \frac{\nabla_x^2}{4m} -\vec{V} (\vec{r}) +i \Gamma \frac{\delta(r)}{4 \pi r} \right] \langle 0| T \Phi(x,\vec{r}) \Phi^\dagger (y,\vec{r}') |0\rangle = i \delta(x-y) \delta(\vec{r}-\vec{r}'),
\end{align}
where $\vec{V}(\vec{r})$ is 
\begin{align}
\vec{V} (\vec{r})=\left(
\begin{array}{ccc}
2\delta m - {\cal S}_{Pot}^{(1)}  && -2i{\cal S}_{Pot}^{(2)}\\
-2i{\cal S}_{Pot}^{(2)} && 0\\
\end{array}
\right).
\end{align}
By defining the radial component of Green function as $\vec{G}^{(E,l)}$, Eq.~(\ref{schwinger}) becomes the following form.
\begin{align}
\label{doukei}
\left[ -E - \frac{1}{mr}\frac{d^2}{dr^2} r + \vec{V} (\vec{r}) -i \Gamma \frac{\delta(r)}{4 \pi r} \right] \vec{G}^{(E,0)}(r,r') = \frac{\delta(r-r')}{r^2}.
\end{align}
To determine $\vec{G}^{(E,0)}$, Eq.~(\ref{doukei}) must be solved in a proper boundary condition.
For this purpose we consider the terms involved $\Gamma$ as in perturbation series
and by using variable transformation $\displaystyle\vec{\mathrm{g}}(r, r') = r r' \vec{G}^{(E,0)}_{ii}(r,r')$. 
The leading order's solution  $\vec{\mathrm{g}}_0(r,r')$ satisfies the equation
\begin{align}
\label{leading}
-\frac{1}{m} \frac{d^2}{dr^2} \vec{\mathrm{g}}_0(r,r') + \vec{V} (\vec{r}) \vec{\mathrm{g}}_0(r,r') - E \vec{\mathrm{g}}_0(r,r') = \delta(r-r').
\end{align}
We get $\vec{\mathrm{g}}_0(r,r')$ in the following form:
\begin{align}
\vec{\mathrm{g}}_0(r,r')=m \vec{\mathrm{g}}_>(r) \vec{\mathrm{g}}_<^{\sf T}(r') \theta (r-r') + m \vec{\mathrm{g}}_<(r) \vec{\mathrm{g}}_>^{\sf T}(r') \theta (r'-r) \hspace{2mm}.
\end{align}
Here The solutions $\vec{\mathrm{g}}_>(r)$ and $\vec{\mathrm{g}}_<(r)$ satisfy the below boundary condition written in \cite{Hisano:2004ds}.
\begin{description}
\item[\mbox{(i)}] $\vec{\mathrm{g}}_{<}(0) = \vec{0}$ , \, $\displaystyle\frac{d~\vec{\mathrm{g}}_{<}(0)}{ d r} = \vec{1}$ ,
\item[\mbox{(ii)}] $\vec{\mathrm{g}}_>(0) = \vec{1}$ , \, $\vec{\mathrm{g}}_>(r)$ has only outgoing
  wave at $r\rightarrow \infty$.
\end{description}
Also, $\vec{\mathrm{g}}_{<(>)}(r)$ satisfies, respectively, the following differential equation 
\begin{align}
\label{schrodinger}
  -\frac{1}{m} \frac{d^2}{dr^2} \vec{\mathrm{g}}_{<(>)}(r) + \vec{V} (\vec{r}) \vec{\mathrm{g}}_{<(>)}(r)
  = E \vec{\mathrm{g}}_{<(>)}(r) \hspace{2mm}.
\end{align}
Furthermore, $\widetilde{\tau}$ does not exist at $r\rightarrow \infty$ when
$E < 2\delta m$, so
\begin{align} [\vec{\mathrm{g}}_{>}(r)]_{ij}
  \raisebox{-2.5mm}{$\left|_{r\rightarrow \infty} \right.
    $}= \delta_{i2} d_{2j}(E) e^{i\sqrt{mE}r} \hspace{2mm}.
\end{align}
When we calculate $\vec{\mathrm{g}}_>(r)$ by using the first-order perturbation, cross section for annihilation channel $f$ can be written as
\begin{align}
\label{cross-section}
\sigma_2^{(S)} v|_f = [\widetilde{\Gamma}_{f} ]_{11}  d_{21}(mv^2/4) d^*_{21}(mv^2/4) \hspace{2mm}.
\end{align}
In addition, when we write the sum of each annihilation channel~$f$ as
$\Gamma=\sum_{f} \widetilde{\Gamma}_f$, total cross section satisfies
 \begin{align}
\sigma^{(S)}_2 v=\Gamma_{11} d_{21}(mv^2/4)  d_{21}^*(mv^2/4) \hspace{2mm}.
\end{align}
Therefore it is necessary to solve the equation about
$\vec{\mathrm{g}}_{>}(r)$ to determine $d_{2j}$ and cross section.

\subsection{DM signature}
Next, we calculate gamma-ray flux from DM annihilation that occurred in
our galactic center. The spectrum of gamma-ray has two types. One of these
is the line gamma-ray spectrum and the other is the continuum gamma-ray
spectrum. Because the DM moves non-relativistically, the line spectrum lies at
the mass of DM. On the other hand, continuum gamma-ray signal comes
from jets from the DM annihilation. For example, produced $\pi$ mesons
from DM annihilation decay into $\gamma \gamma$. Such a signal is useful
when the cosmic background is well known.

The gamma-ray flux from DM annihilation used by indirect detection experiments \cite{Bergstrom:1997fj,Ackermann:2015zua} is given by
\begin{align}
  \frac{d\Phi_{\gamma}}{d E} = \frac{1}{4\pi} \frac{1}{m_{\chi}^2} \sum_f \frac{d N_f}{d E}
  \frac{\langle \sigma v \rangle_f}{2}  \times J ,
\end{align}
where $m_\chi$ is DM mass, and $dN_f/dE \cdot dE$ is the numbers of photon
derived from annihilation channel $f$ whose energy is between $E$ and
$E+dE$. $\langle \sigma v \rangle$ is the DM annihilation cross section
averaged with the velocity.

$J$ is called ``$J$-factor'' which is determined by an astrophysical parameter and is given as  
\begin{align}
J=\int_{\rm line~of~sight} dl(\theta) \int_{\Delta \Omega} d \Omega \  \rho^2,
\end{align}
where $\Delta \Omega$ is the angular resolution and $\rho$ is DM
density in our galaxy. N-body simulations show some DM halo
profiles. For example, NFW \cite{Navarro:1996gj}, Burkert \cite{Burkert:1995yz}, and Einasto~\cite{Graham:2005xx,Navarro:2008kc} profiles are widely used.
\begin{align}
&\rho_{\rm NFW} = \frac{\rho_s}{(r/r_s)(1+r/r_s)^2} \hspace{2mm}, \\[2mm]
&\rho_{\rm Burkert} = \frac{\rho_s}{(1+r/r_s)(1+(r/r_s)^2)}\hspace{2mm}, \\[2mm]
&\rho_{\rm Einasto} = \rho_s\exp \left\{ -\frac{2}{\alpha} \left[ \left( \frac{r}{r_s}\right)^\alpha-1 \right]\right\}.
\end{align}
Here $\rho_s$, $r_s$, and $\alpha$ are determined from
observations~\cite{Cirelli:2010xx}.
Next, we discuss quantities determined from particle physics. Cross
section $\langle \sigma v \rangle$ is determined from
Eq.~(\ref{cross-section}) as mentioned above. We calculate energy
spectrum $dN_f/dE$ with pythia~\cite{Sjostrand:2014zea} for each
annihilation channels.

For the case $Z$-boson annihilate to gamma is shown in Fig.~3.
\begin{figure}[H]
\begin{center}
\includegraphics[scale=0.6]{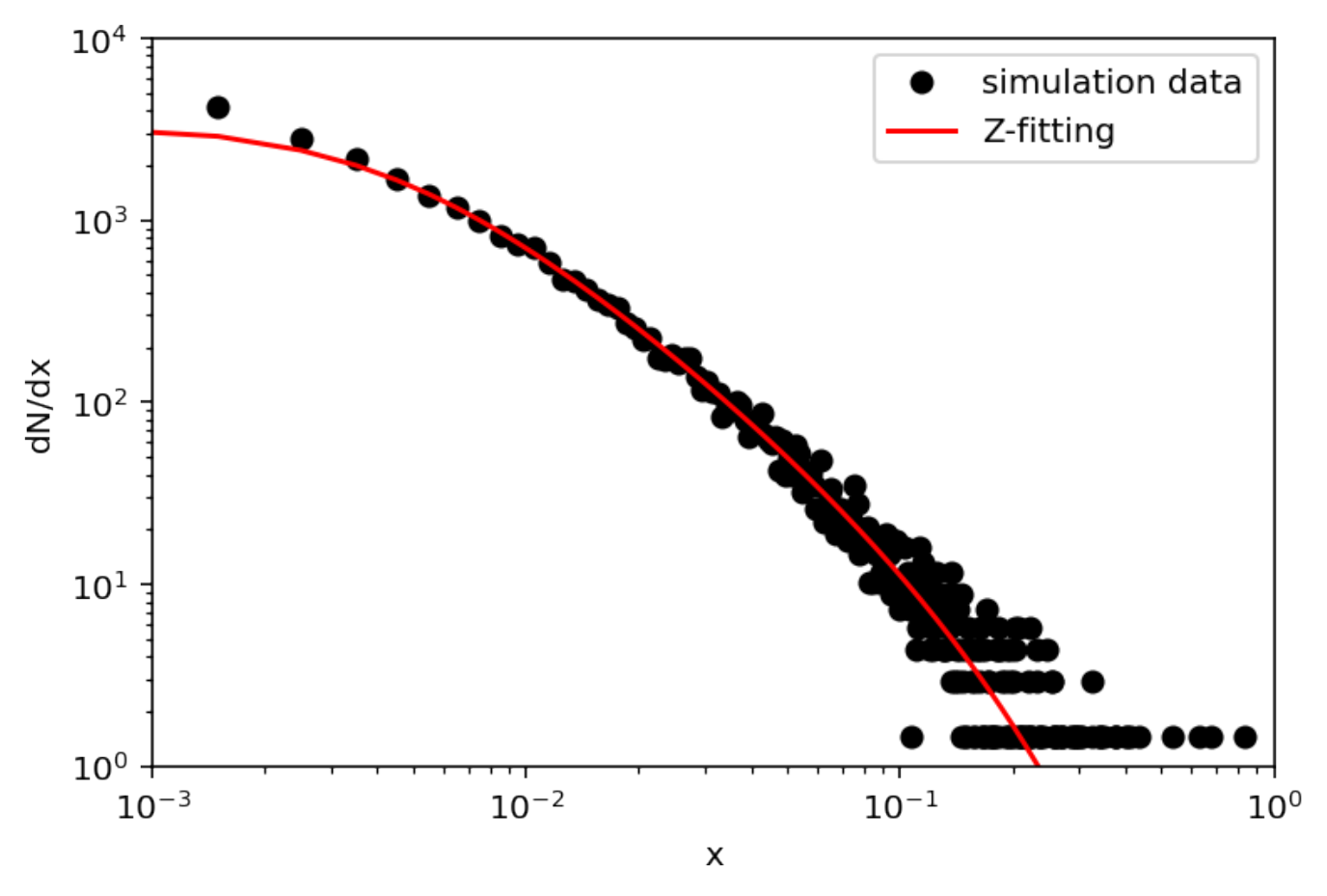}
\end{center}
\caption{Energy spectrum in the case where DM decays into $Z$-boson.}
\label{zboson}
\end{figure}
\begin{equation}
\frac{d N_Z}{d x} = \frac{0.642e^{-7.98x}}{x^{1.59} + 1.88\times10^{-4}},
\end{equation}
where $x=E/m_\chi$.

For the case $W$-boson and $\tau$ annihilate to gamma is respectively shown in Fig.~4.
\begin{figure}[H]
\begin{center}
\begin{tabular}{c}
\begin{minipage}{0.5\hsize}
\begin{center}
\includegraphics[bb=0 0 0 0,scale=0.6,viewport=0 0 640 284,clip]{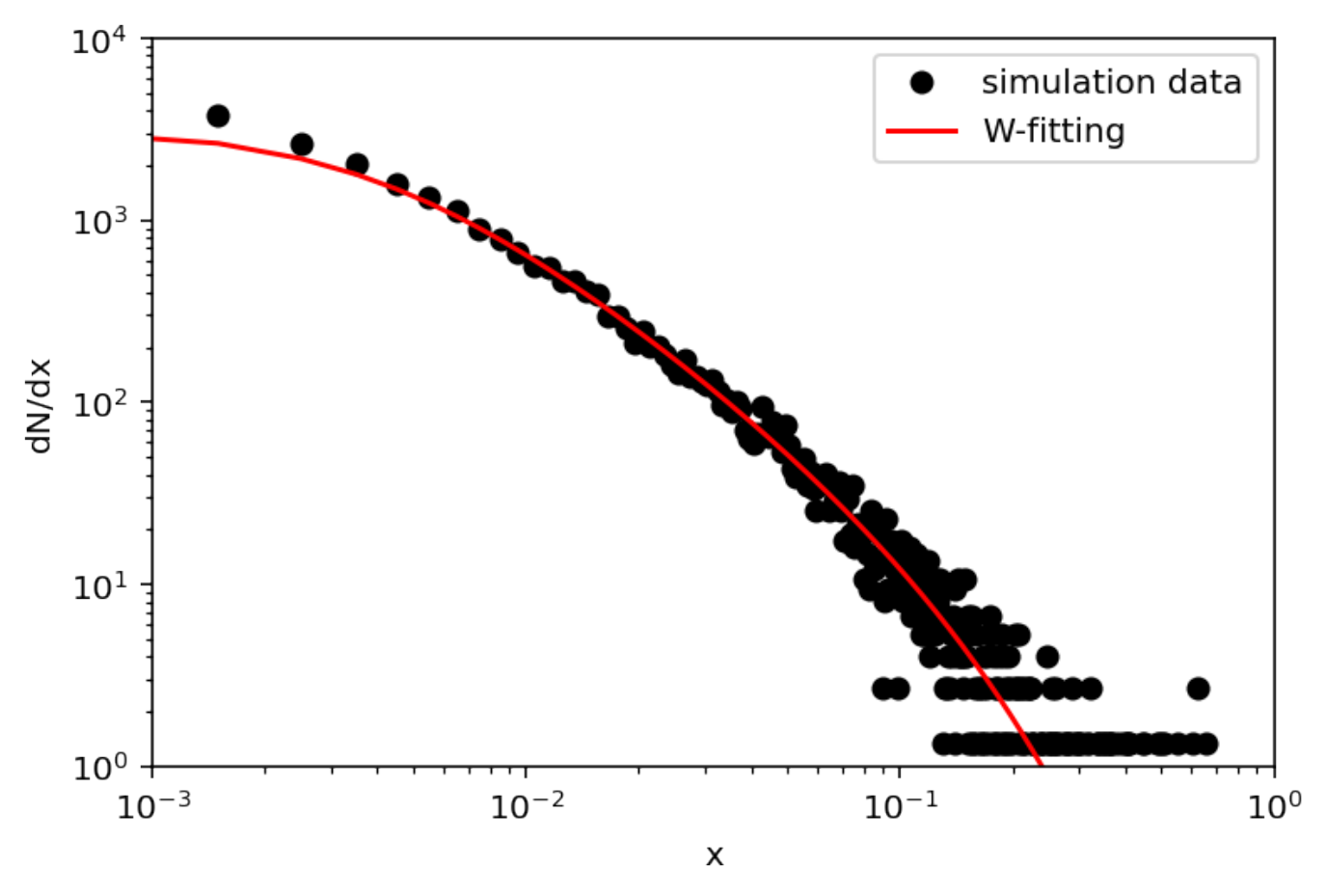}
\end{center}
\end{minipage}
\begin{minipage}{0.5\hsize}
\begin{center}
\includegraphics[bb=0 0 0 0,scale=0.6,viewport=0 0 640 284,clip]{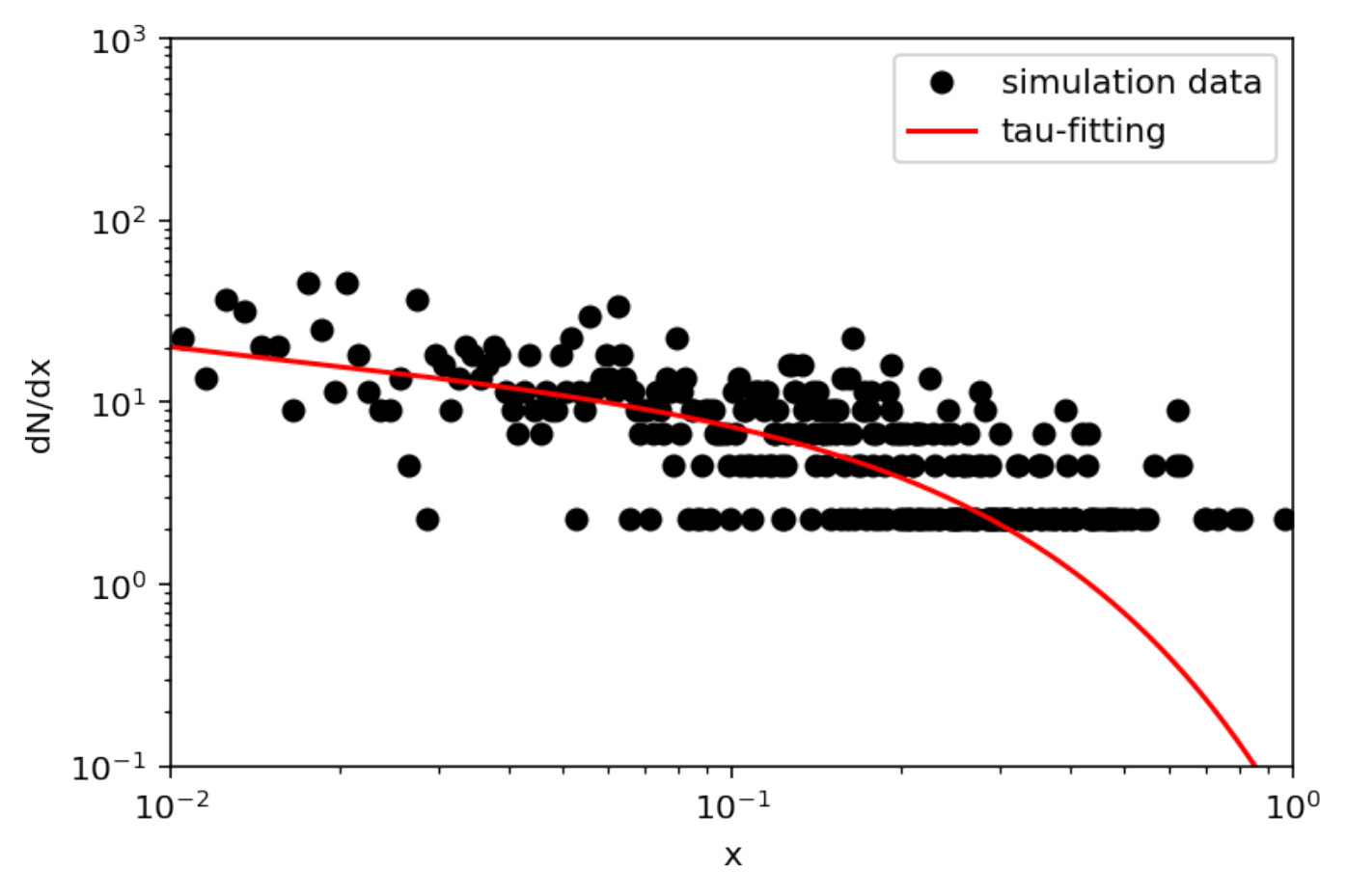}
\end{center}
\end{minipage}\\
\end{tabular}
\caption{The left graph corresponds to the case where DM decays into
  $W$-boson. The right graph corresponds to the case where DM decays into
  $\tau$. }
\end{center}
\label{Wtau}
\end{figure}
\begin{align}
\frac{d N_W}{d x} =  \frac{ 0.931e^{-8.60x}}{x^{1.49} + 2.87\times10^{-4}} ,\quad 
\frac{d N_\tau}{d x} = \frac{0.0264e^{-5.24x}}{x^{3.89\times10^{-4}} -0.997}.
\end{align}
For directly decaying into $\gamma \gamma$ case, the spectrum becomes
\begin{equation}
\frac{d N_\gamma}{d E} =  2\delta(E-m).
\end{equation}

\section{Result}
In this section, we will discuss our results. First, we find that the
peak of cross section appears with the fixed value of $\delta m$ by
using the parameters which are presented in the following tables for
numerical calculations~\cite{Kubo:2018xrk}: The dimensionless
parameters are shown in Table~1 and the dimensionful parameters are
shown in {\rm GeV} units in Table~2 and the mixing parameters are
shown in Table~3, Table~4 and Table~5.  These parameters are
satisfied with a positive solution for Li problem in cosmology.  We
should keep in our mind the fact that stau mass is limited up to about
430~${\rm GeV}$ by ATLAS experiment~\cite{Aaboud:2019trc}. Thus we
have to set up the mass parameters of neutralino and stau to adjust
peak position at near 430~${\rm GeV}$. We assume, however, that the
calculation results are not affected if the mass values are
slightly different from the values in the tables.

\begin{table}[H]
\begin{center}
  \begin{tabular}{|c||c|c|c|c|c|c|c|c|}\hline
    parameter &  $g$  &  $g'$ & $s_w^2$ &  $\tan\beta$ & $\cot\alpha$ & $y_1{^1}$ & $y_2{^2}$ &$y_3{^3}$ \\\hline
    value &  0.6387 &  0.3623 & 0.23 &  24.21 & -24.21& $1.0 \times 10^{-5}$ & $1.5 \times 10^{-2}$ & 0.2542\\ \hline
\end{tabular}
\caption{Dimensionless parameters for numerical calculations~\cite{Kubo:2018xrk} are presented.}
\end{center}
\end{table}
\begin{table}[H]
  \begin{flushleft}
  \begin{tabular}{|c||c|c|c|c|c||c|c|c|c|c|c|c|}\hline
    parameter & $m$ & $m_{\widetilde{{\tau}}}$ &  $m_{C_1}$ & $\mu$ & $A_0$ & $m_e {\scriptscriptstyle\times10^{-3}}$ & $m_\mu$ &$m_\tau$ & $v$ & $m_w$& $m_z$
    & $m_{h^0}$ \\ \hline
    value &  379.596576 &  379.606567 & 725.76 & 1.776 & -3098.1& $0.511$   & 0.105 & 1.776  & 243.5786 & 80.2  & 91.19& 125.18 \\\hline  
\end{tabular}
\caption{The mass parameters used for numerical calculations are
  presented in {\rm GeV} unit~\cite{Kubo:2018xrk}.}
\end{flushleft}
\end{table}

\begin{table}[H]
\begin{center}
\scalebox{0.8}[0.8]{
\begin{tabular}{c|cc||c|cc}
\hline \hline
\multicolumn{6}{c}{chargino} \\ \hline \hline
\multicolumn{3}{c||}{real part } & \multicolumn{3}{c}{imaginary part } \\ \hline \hline 
${\rm Re}U{_i}^j$ & $j=1$ & $j=2$ & ${\rm Im}U{_i}^j$ & $j=1$ & $j=2$  \\ \hline
$i=1$ & $-9.97043970\times 10^{-1}$ & $7.68330827\times 10^{-2}$ & $i=1$ & $-2.92376714\times 10^{-16}$ & $9.05483611\times 10^{-18}$   \\ \hline
$i=2$ & $7.68330827\times 10^{-2}$ & $9.97043970\times 10^{-1}$ & $i=2$ & $-4.61092153\times 10^{-17}$ & $-7.73222256\times 10^{-16}$ \\  \hline \hline 
${\rm Re}V{_i}^j$ & $j=1$ & $j=2$ & ${\rm Im}V{_i}^j$ & $j=1$ & $j=2$  \\ \hline
$i=1$ & $-9.99396601\times 10^{-1}$ & $3.47337437\times 10^{-2}$ & $i=1$ & $7.74343187\times 10^{-16}$ & $0.00000000\times 10^{0}$  \\ \hline
$i=2$ & $3.47337437\times 10^{-2}$ & $9.99396601\times 10^{-1}$  & $i=2$ & $0.00000000\times 10^{0}$ & $7.74343187\times 10^{-16}$\\ 
\hline
\end{tabular}}
\caption{The values of the unitary matrix that diagonalizes chargino.}
\label{tab:chargino}
\end{center}
\end{table}

\begin{table}[H]
\begin{center}
\scalebox{0.9}[0.9]{
\begin{tabular}{c||cccc}  
\hline \hline
\multicolumn{5}{c}{neutralino} \\
\hline \hline
\multicolumn{5}{c}{real part} \\ \hline
${\rm Re}N_{\widetilde{G}^a}^b$ & $b=1$& $b=2$ & $b=3$ & $b=4$ \\ \hline
$a=1$ & $9.99599220\times 10^{-1}$ & $-2.13656993\times 10^{-3}$ & $2.72724410\times 10^{-2}$ & $-7.28340424\times 10^{-3}$  \\
$a=2$ & $3.78724402\times 10^{-3}$ & $9.98231872\times 10^{-1}$ & $-5.41258402\times 10^{-2}$ & $2.42730150\times 10^{-2}$  \\
$a=3$ & $-2.23811821\times 10^{-17}$ & $4.82387775\times 10^{-17}$ & $6.11227032\times 10^{-16}$ & $-6.11568206\times 10^{-16}$ \\
$a=4$ & $2.42682988\times 10^{-2}$ & $-5.55057268\times 10^{-2}$ &  $-7.05176402\times 10^{-1}$ & $7.06439245\times 10^{-1}$  \\ \hline \hline
\multicolumn{5}{c}{imaginary part} \\ \hline \hline
${\rm Im}N_{\widetilde{G}}{_a}^b$ & $b=1$ & $b=2$ & $b=3$ & $b=4$  \\ \hline
$a=1$ & $-5.13227991\times 10^{-16}$ & $2.88642911\times 10^{-19}$ & $-1.03992019\times 10^{-17}$ & $-1.74565311\times 10^{-18}$  \\
$a=2$ & $-1.78170707\times 10^{-18}$ & $-2.33972913\times 10^{-16}$ & $5.73625399\times 10^{-18}$ & $1.43366353\times 10^{-18}$  \\
$a=3$ & $1.40749921\times 10^{-2}$ & $-2.11584124\times 10^{-2}$ & $-7.06436727\times 10^{-1}$ & $-7.07319847\times 10^{-1}$ \\
$a=4$ & $5.84094614\times 10^{-18}$ & $-8.37388039\times 10^{-18}$ &  $-6.12074422\times 10^{-16}$ & $-6.13058014\times 10^{-16}$  \\
\hline
\end{tabular}}
\caption{The values of the diagonalizing unitary matrix of neutralino.}
\end{center}
\end{table}

\begin{table}[H]
\begin{center}
\scalebox{0.7}[0.7]{
\begin{tabular}{c||cccccc}  
\hline \hline
\multicolumn{7}{c}{slepton} \\
\hline \hline
\multicolumn{7}{c}{real part} \\ \hline
${\rm Re}N_{\widetilde{l}}{_A}^B$ & $B=1$ & $B=2 $& $B=3 $ & $B=4$ & $B=5$ & $B=6$ \\ \hline
$A=1$ & -6.31168044$\times 10^{-9}$ & $2.63719358\times 10^{-6}$ & $-1.71009221\times 10^{-1}$ & $6.25408690\times 10^{-11}$ &$-5.57777646\times 10^{-8}$ & $-9.85269428\times 10^{-1}$ \\
$A=2$ & $-4.88052654\times 10^{-6}$ & $-1.47061845\times 10^{-2}$ & $1.26260288\times 10^{-5}$ & $-2.05256544\times 10^{-8}$ & $-7.02536796\times 10^{-1}$ & $-2.18090514\times 10^{-6}$ \\
$A=3$ & $-1.68078331\times 10^{-5}$ & $-2.88036700\times 10^{-8}$ & $-1.53915085\times 10^{-10}$ & $ -1.65361745\times 10^{-1}$ & $2.67385497\times 10^{-8}$ & $-4.95673580\times 10^{-11}$ \\
$A=4$ & $1.00627884\times 10^{-7}$ & $-5.03237436\times 10^{-5}$ &  $9.85269426\times 10^{-1}$ & $-3.17467808\times 10^{-10}$ & $1.00201302\times 10^{-5}$ & $-1.71009221\times 10^{-1}$ \\ 
$A=5$ & $6.62240877\times 10^{-2}$ & $6.99028792\times 10^{-1}$ &  $7.13839275\times 10^{-5}$ & $-6.76095272\times 10^{-6}$ & $-1.46315410\times 10^{-2}$ & $-8.58412190\times 10^{-6}$ \\
$A=6$ & $-8.24980621\times 10^{-1}$ & $5.55284568\times 10^{-2}$ &  $5.85237100\times 10^{-6}$ & $8.39260849\times 10^{-5}$ & $-1.15816598\times 10^{-3}$ & $-7.02385695\times 10^{-7}$  \\ \hline \hline
\multicolumn{7}{c}{imaginary part} \\ \hline \hline
${\rm Im}N_{\widetilde{l}}{_A}^B$ & $B=1$ & $B=2$ & $B=3$ & $B=4$ & $B=5$ & $B=6$ \\ \hline
$A=1$ & $-6.69074517\times 10^{-9}$ & $2.68186615\times 10^{-6}$ & $-3.36466824\times 10^{-15}$ & $-6.50682427\times 10^{-11}$ &$-7.01708566\times 10^{-8}$ & $4.44552436\times 10^{-23}$ \\
$A=2$ & $3.55520850\times 10^{-6}$ & $-1.48888937\times 10^{-2}$ & $-5.52344800\times 10^{-8}$ & $1.51693448\times 10^{-8}$ & $-7.11339651\times 10^{-1}$ & $6.71707263\times 10^{-16}$ \\
$A=3$ & $1.00325957\times 10^{-4}$ & $-4.18395754\times 10^{-10}$ & $2.93857601\times 10^{-10}$ & $9.86232976\times 10^{-1}$ & $6.16275932\times 10^{-10}$ & $1.39096557\times 10^{-17}$ \\
$A=4$ & $1.44670207\times 10^{-7}$ & $-5.11527461\times 10^{-5}$ &  $1.15540529\times 10^{-12}$ & $7.71687277\times 10^{-11}$ & $1.02274109\times 10^{-5}$ & $-9.52745431\times 10^{-21}$ \\ 
$A=5$ & $-4.78727158\times 10^{-2}$ & $7.10102998\times 10^{-1}$ &  $-8.62478816\times 10^{-9}$ & $4.88605503\times 10^{-6}$ & $-1.48648974\times 10^{-2}$ & $1.39571889\times 10^{-17}$ \\
$A=6$ & $5.59222261\times 10^{-1}$ & $5.99258994\times 10^{-2}$ &  $3.23933295\times 10^{-10}$ & $-5.68738081\times 10^{-5}$ & $-1.24999822\times 10^{-3}$ & $1.05557136\times 10^{-18}$ \\
\hline
\end{tabular}}
\caption{The values of the real part and imaginary part of diagonalizing unitary matrix of slepton.}
\end{center}
\end{table}

We calculate for several annihilation channels. \mbox{(1)} we show the
result of the cross sections DM decaying directly into two gammas for
the corresponding $\delta m$ as displayed in Fig.~5.
\begin{figure}[H]
\label{each-dm}
\begin{center}
\includegraphics[scale= 0.85]{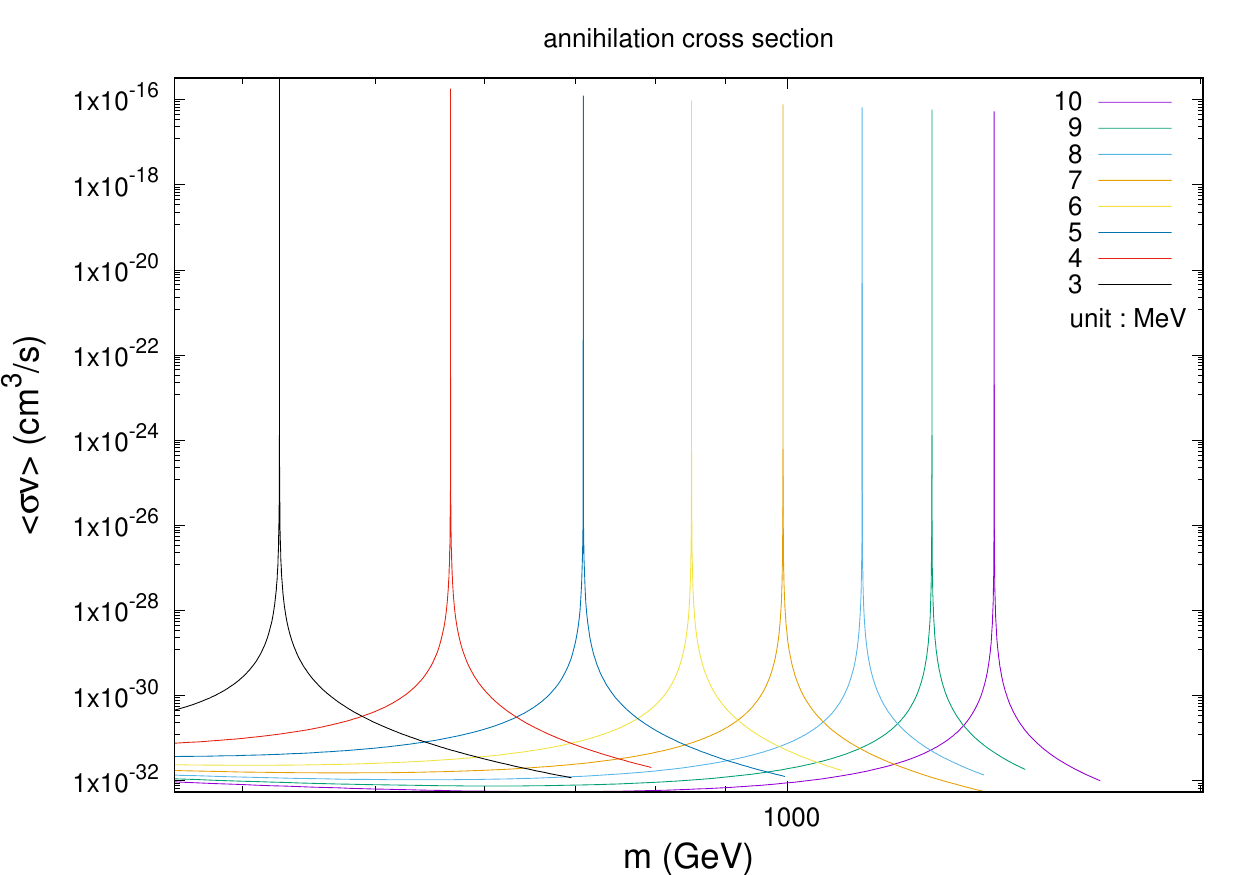}
\vskip8mm
\end{center}
\caption{Annihilation cross section to photons per $\delta m$. Each
  graph name represents $\delta m$ in ${\rm MeV}$ units. }
\end{figure} {}From Fig.~5 we realize that annihilation cross
sections reach the height point at different DM mass, and the cross sections
decrease as its $\delta m$ increases. \mbox{(2)} cross section of DM
decaying directly into $Z$-boson, $W$-boson, and $\tau$ are shown in
Fig.~6 and Fig.~7, respectively.
\begin{figure}[H]
\begin{center}
\begin{tabular}{c}
\begin{minipage}{0.5\hsize}
\begin{center}
\includegraphics[bb=0 0 0 0,scale=0.6,viewport=0 0 640 284,clip]{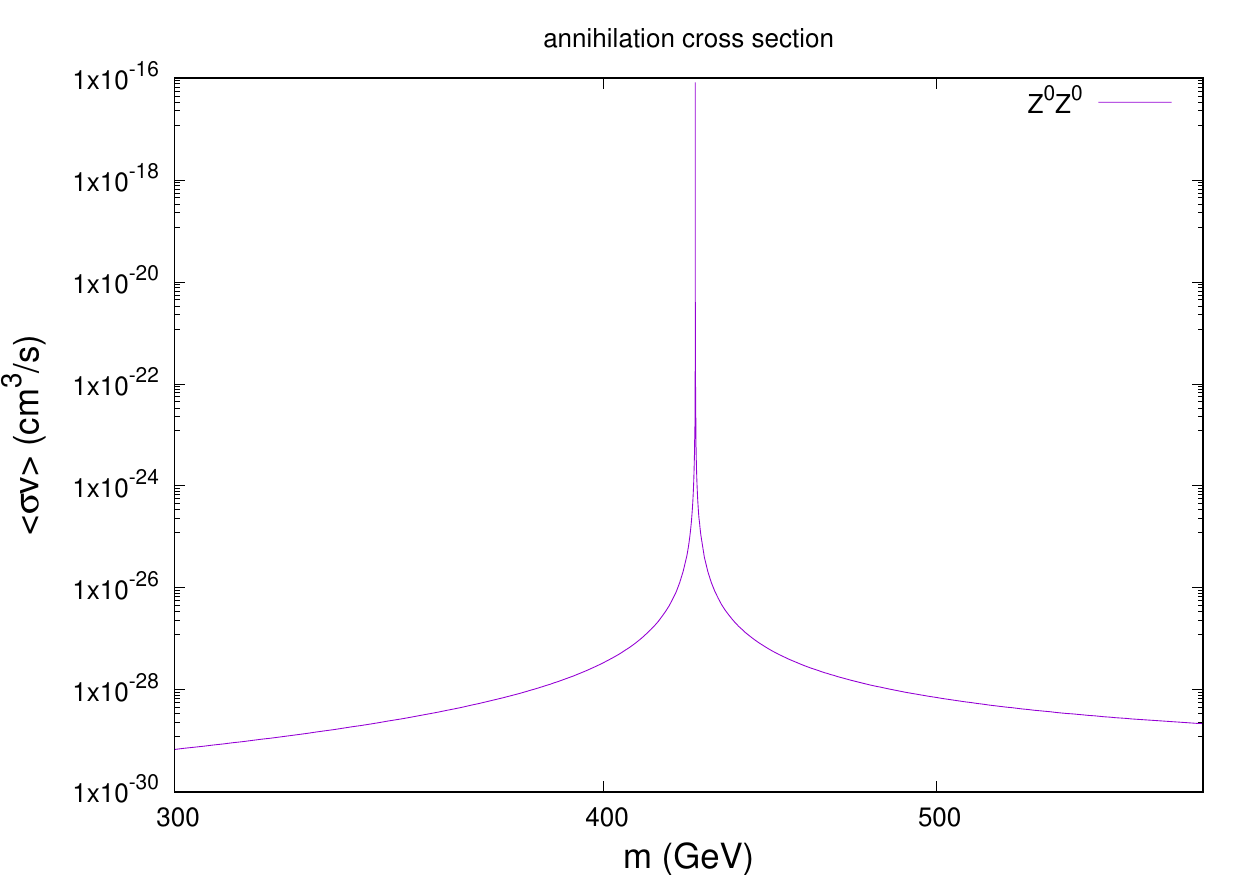}
\end{center}
\end{minipage}
  \begin{minipage}{0.5\hsize}
\begin{center}
\includegraphics[bb=0 0 0 0,scale=0.6,viewport=0 0 640 284,clip]{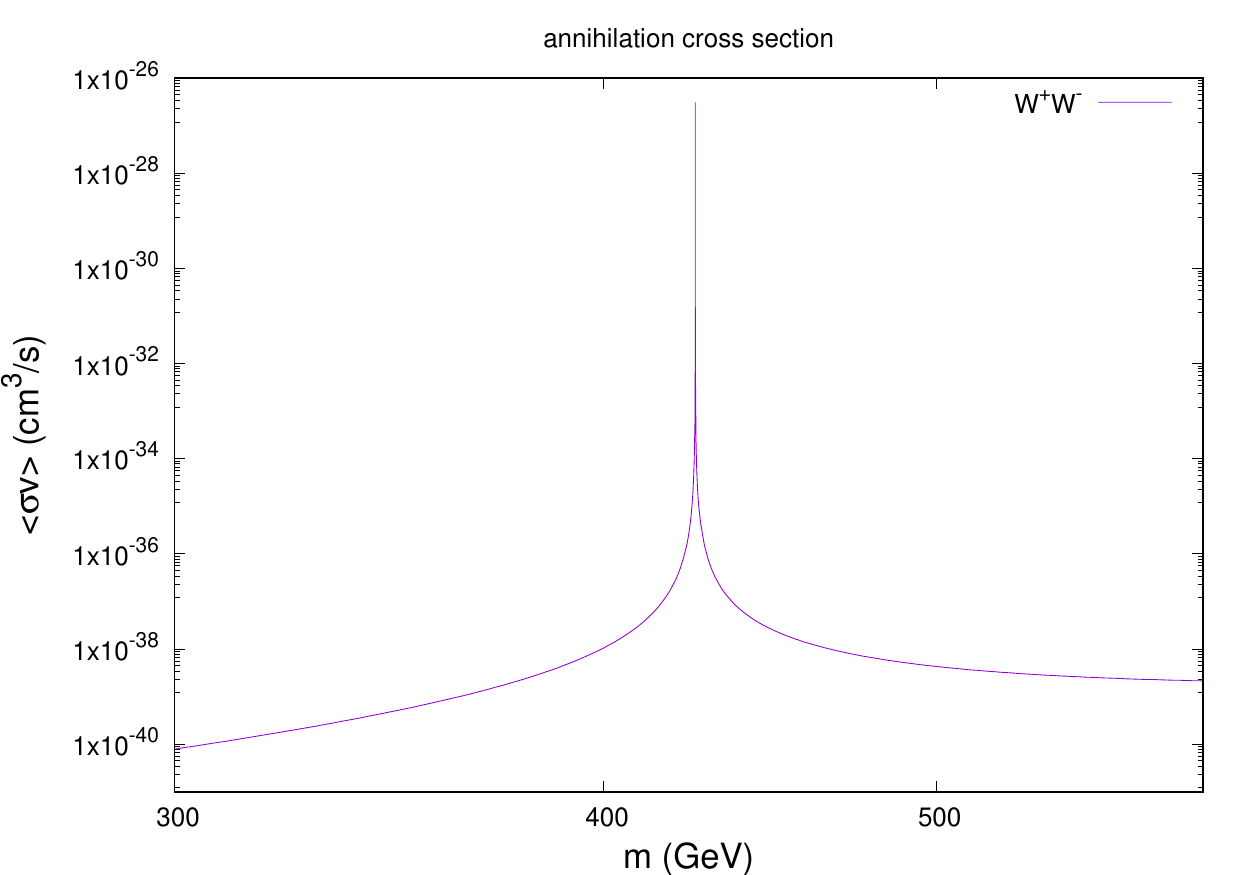}
\end{center}
\end{minipage}
\end{tabular}
\end{center}
\caption{The left(right) plot correspond to annihilating
  $Z^0Z^0$($W^+ W^-$) case, where we set $\delta m=3~{\rm MeV}$. }
\end{figure}
\begin{center}
\begin{figure}[H]
\begin{center}
\includegraphics[scale=0.6]{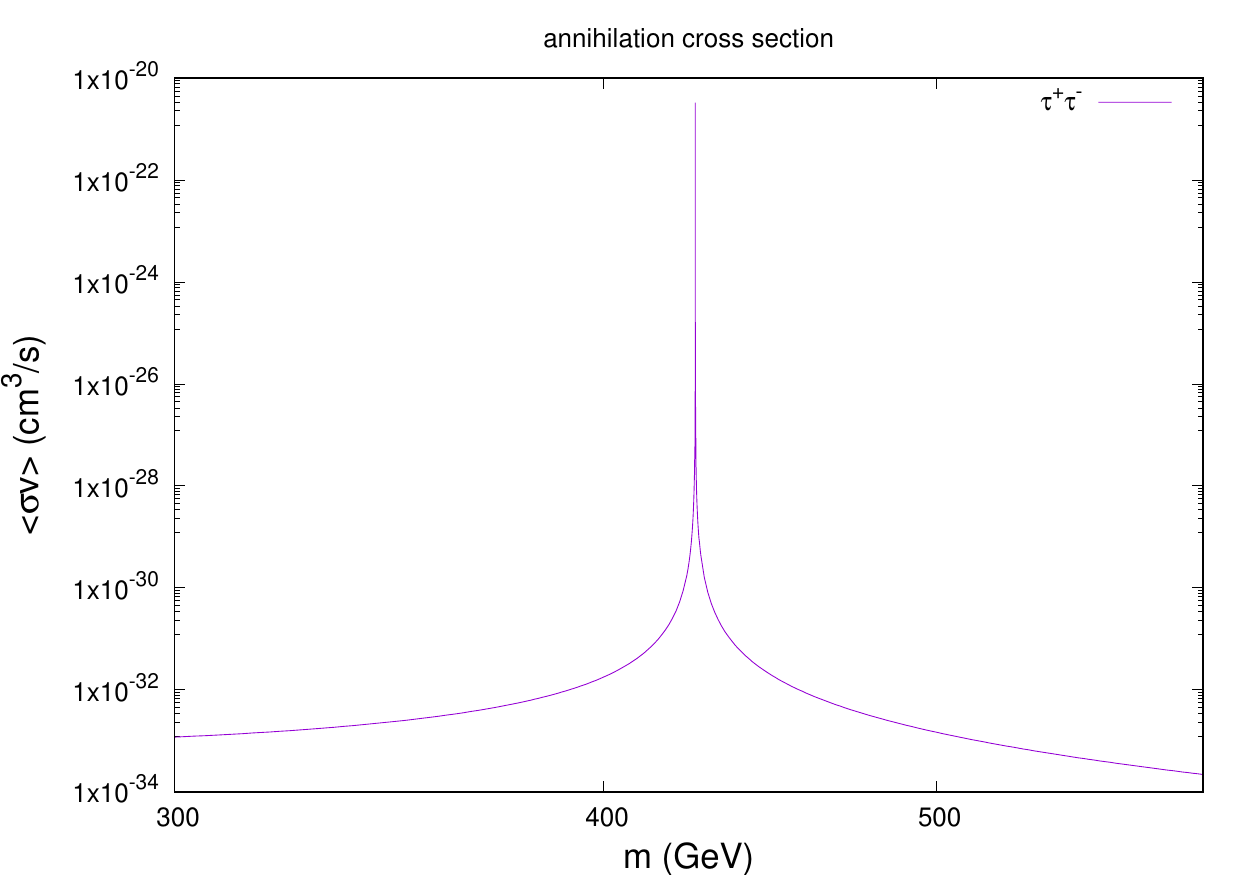}
\end{center}
\caption{Annihilation cross section to $\tau$. Here we set
  $\delta m = 3~{\rm MeV}$.}
\end{figure}
\end{center}
%
As an example of continuum flux, the case where the DM decays into
$Z^0Z^0$, $W^+W^-$ is shown in Fig.~8. Here $J$-factor and angular resolution
are referred from~\cite{Abdallah:2018qtu}.
\begin{figure}[H]
\centering
\includegraphics[scale=0.5]{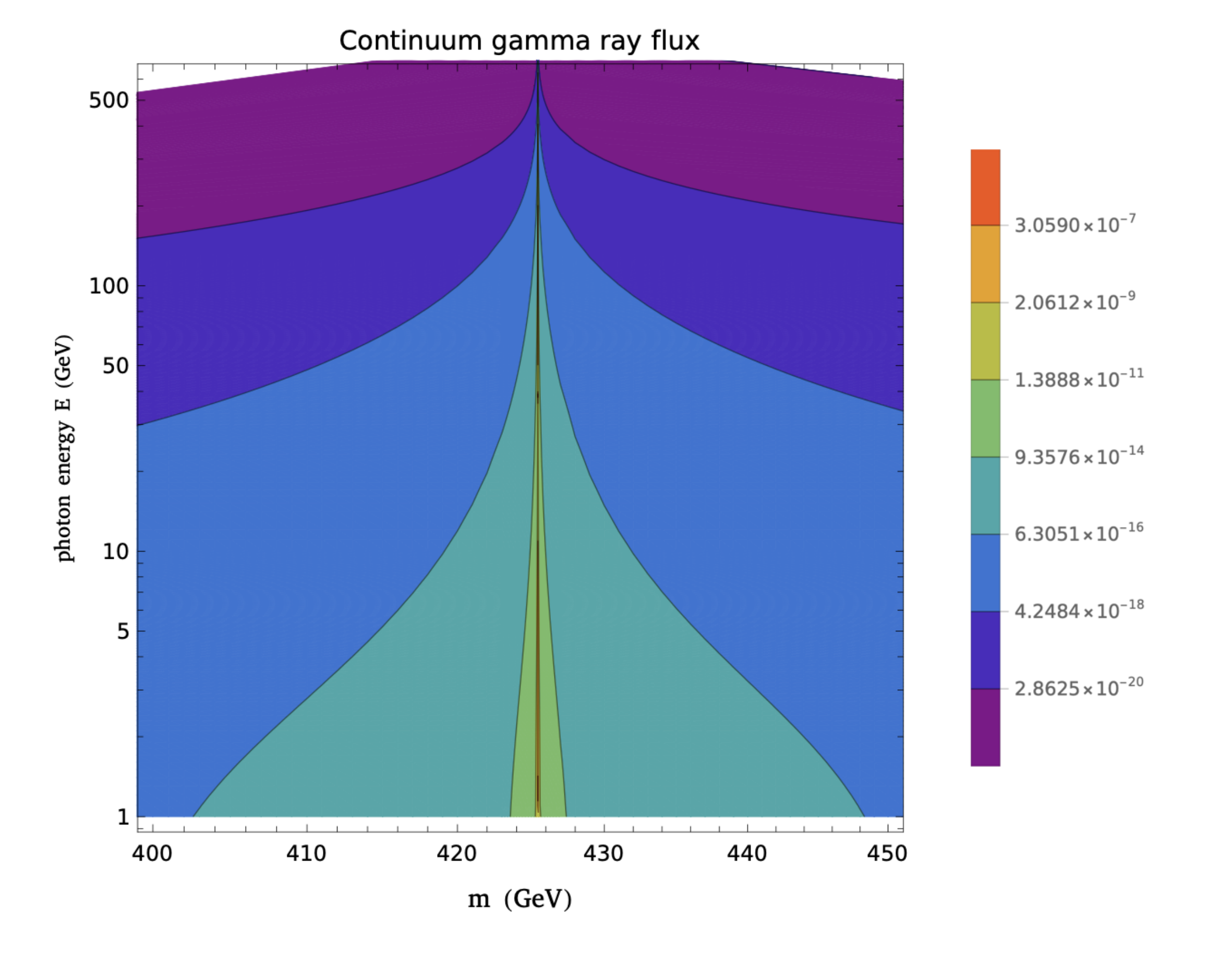}
\caption{Continuum gamma-ray flux in the unit of
  (${\rm cm^{-2} sec^{-1} GeV^{-1}}$) when DM decays into $Z^0Z^0$,
  $W^+W^-$. The color of the graph describes the size of flux. Here, it is
  set to $\delta m = 3~{\rm MeV}$,
  $J=1\times10^{20}~{\rm GeV^2/cm^5}$,
  $\Delta \Omega= 1\times10^{-5}$.}
\end{figure}
In Fig.~8, the horizontal axis represents DM mass and the vertical
axis represents observed photon energy. The color of the figure means the
size of flux which is larger at the peak position of the cross section
and low photon energy and becomes smaller as the distance
increases.

We compare our result to HESS experimental data of
$\chi \chi\rightarrow \gamma \gamma$ focusing on coannihilation
region.  The graph represented this is shown in Fig.~9. As we can see
from this figure cross section invades the prohibited area, therefore
we can constrain the parameter in our model. We also show the
sensitivity projected by CTA~\cite{Hryczuk:2019nql}.  Using this
restricted parameter, we are able to constrain the parameters even
more than now.
\begin{figure}[H]
\begin{center}
\includegraphics{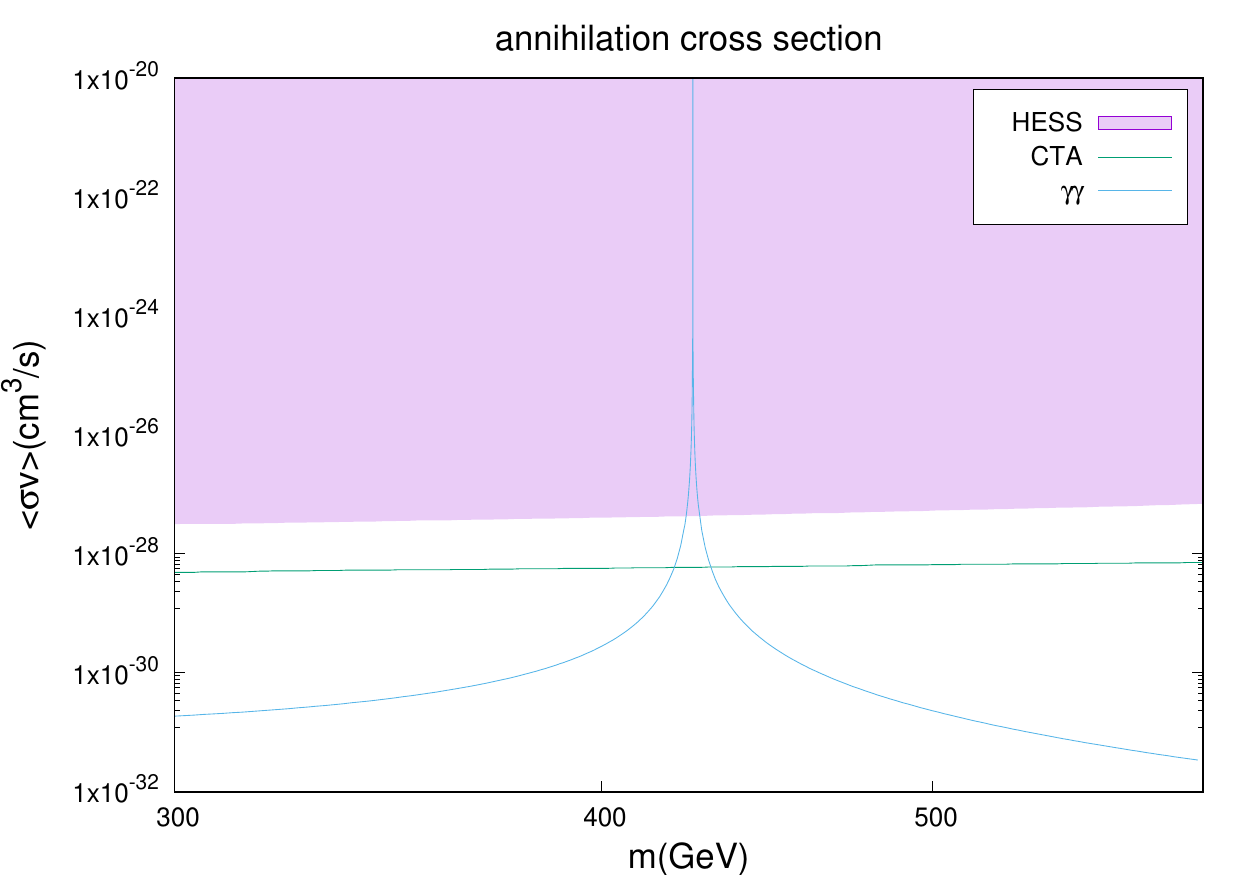}
\vskip10mm
\end{center}
\caption{A comparison of the cross section for $\delta m = 3~{\rm MeV}$
  in the coannihilation region with the HESS
  result~\cite{Abdallah:2018qtu} and projected CTA sensitivity~\cite{Hryczuk:2019nql}. The blue-solid line shows the calculation result, and the purple region shows the HESS result. The green-solid line shows the CTA sensitivity.}
  \label{hess}
\end{figure}

Next, we discuss the result of comparing DMs annihilation to $\chi \chi \rightarrow \tau^+ \tau^-$ channel with HESS and Fermi-LAT experimental data is shown in Fig.~10.
\begin{figure}[H]
\begin{center}
\includegraphics{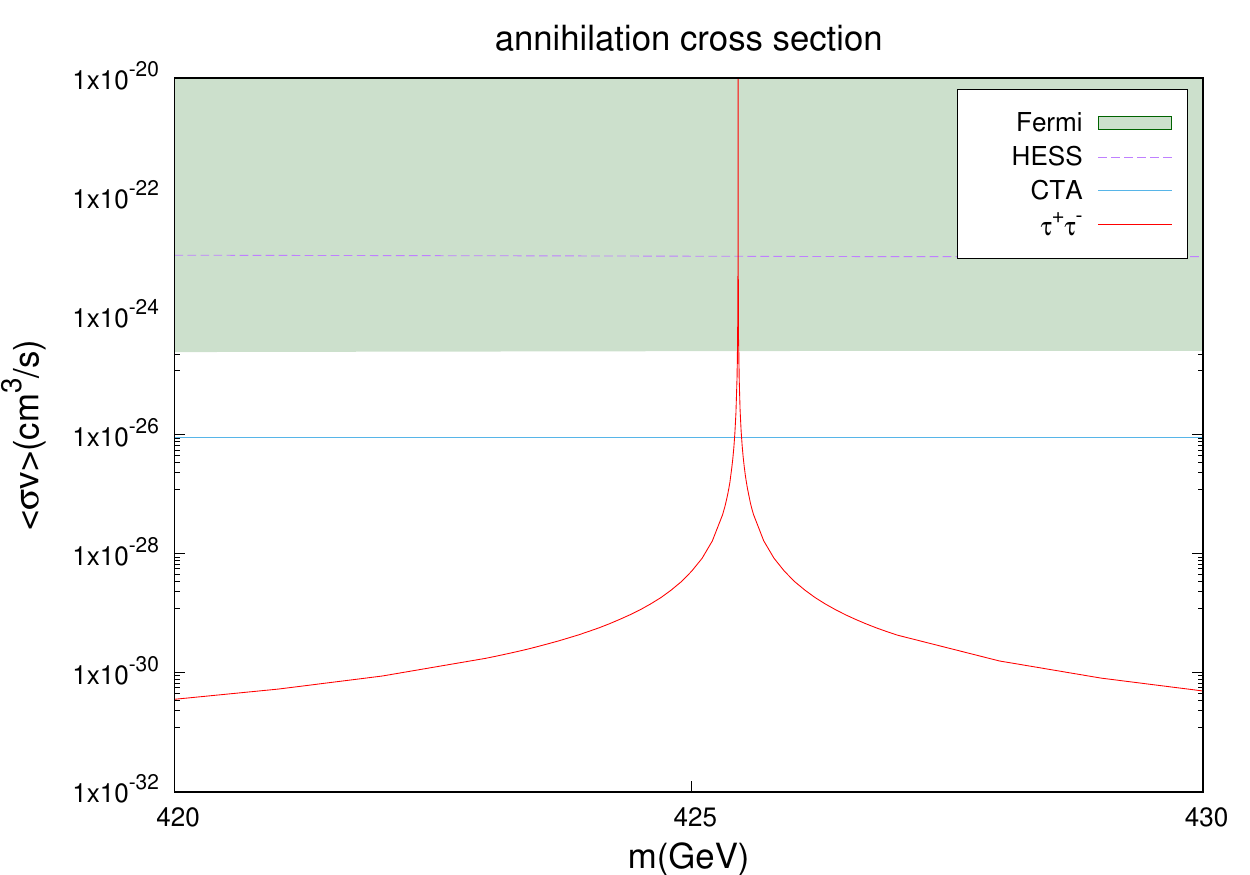}
\vskip10mm
\end{center}
\caption{A comparison of the cross section for
  $\delta m = 3~{\rm MeV}$ for $\tau^+\tau^-$ channel in the
  coannihilation region with the HESS
  result~\cite{Abramowski:2014tra}, Fermi-LAT result~\cite{Ackermann:2015zua}, and projected CTA sensitivity~\cite{Hryczuk:2019nql}. The red-solid line shows the
  calculation result, and the purple dotted-line shows the HESS result. 
  The green area describes the upper limit by Fermi-LAT, and blue-solid line shows CTA sensitivity~\cite{Hryczuk:2019nql}.}
\label{}
\end{figure}
We also draw the projected CTA sensitivity line same as of $\chi \chi \rightarrow \gamma \gamma$ case. Clearly, we cannot limit in most parameter region for $\tau$ channel case, even
for other channels. Thus we put the limitation of parameter region by using
$\chi \chi \rightarrow \gamma \gamma$ channel.

By applying the above-mentioned method to other $\delta m$ cases as
well, we can limit the parameter between $\delta m$ and DM mass. The
restricted parameter region is drawn in Fig.~11. In this figure, the blue-area describes the constrained region of our model.
\begin{figure}[H]
\begin{center}
\includegraphics[scale=0.6]{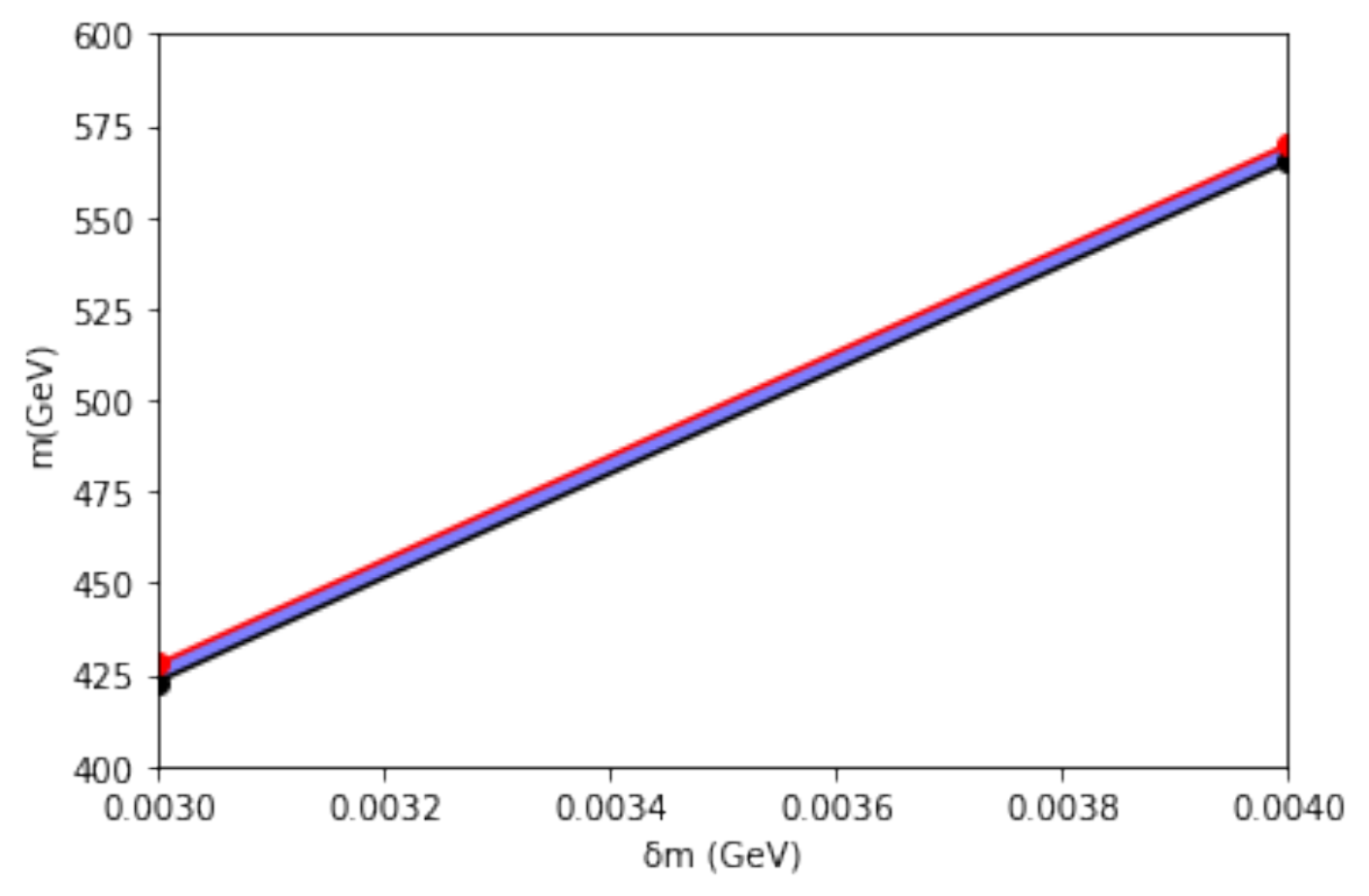}
\end{center}
\caption{The parameter region is limited by HESS experimental data. Parameters in colored areas are restricted.}
\end{figure}

We see that the narrow band in Fig.~11, which presents the limit
from the current experimental restrictions on $\delta m$. In
Fig.~12. we show that the future planned sensitivity of CTA that
restricts about 100 times stronger than the present limit. It is found
that the limit for $\delta m$ can be set in a wide range in the
future.
\begin{figure}[H]
\begin{center}
\begin{tabular}{c}
\begin{minipage}{0.5\hsize}
\begin{center}
\includegraphics[bb=0 0 0 0,scale=0.6,viewport=0 0 640 284,clip]{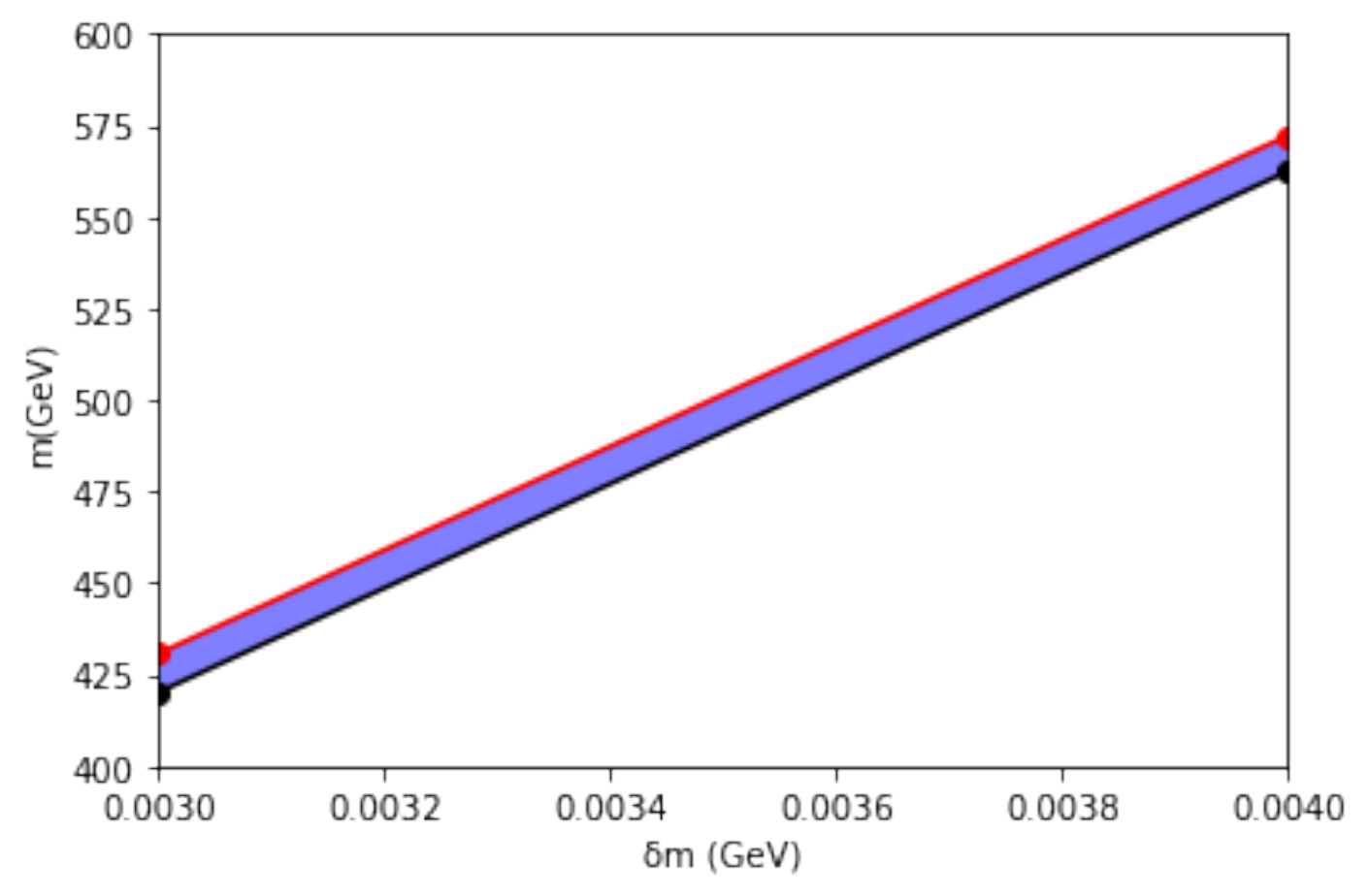}
\end{center}
\end{minipage}

\begin{minipage}{0.5\hsize}
\begin{center}
\includegraphics[bb=0 0 0 0,scale=0.6,viewport=0 0 640 284,clip]{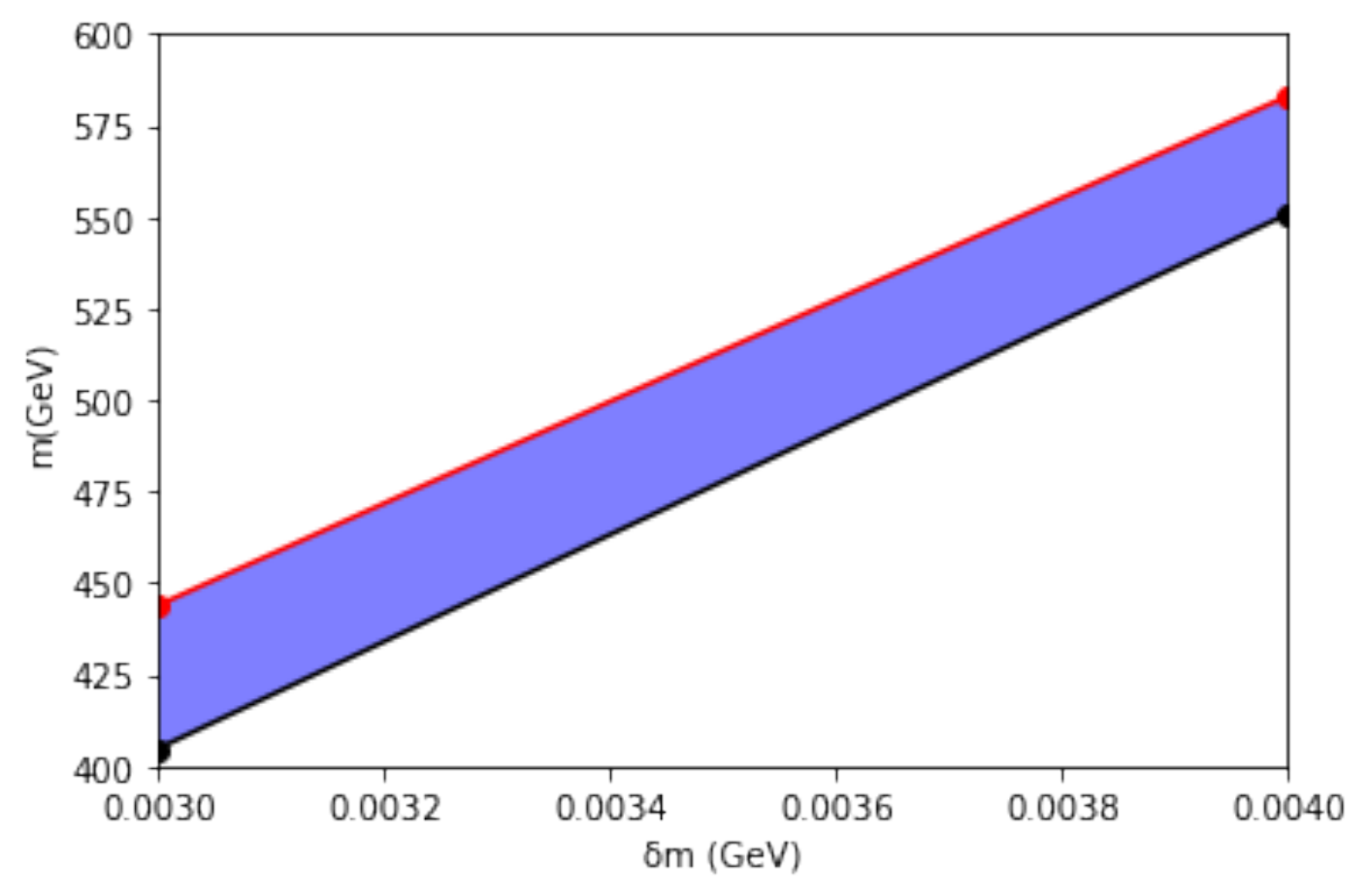}\end{center}
\end{minipage}\\

\end{tabular}
\caption{The left graph corresponds to the limited parameter region constrained by projected CTA sensitivity. The right graph corresponds to the case where the limit
  is 100 times stronger.  As experimental limits increase, more areas
  are restricted.}
\end{center}
\end{figure}

\section{Conclusion}

We have investigated neutralino dark matter in the framework of
MSSM. Hereby the mechanism of Sommerfeld enhancement is taken into
account for the calculation of dark matter annihilation cross section
and flux in the coannihilation region. In this region, if lepton
flavor is violated, Lithium problem in cosmology is solved.
The cross sections for several annihilation channels are shown in
Fig.~5, Fig.~6 and Fig.~7: The dependence of $\delta m$ on cross
sections of ${\chi} \widetilde{\chi} \rightarrow \gamma \gamma$ channel
is shown in Fig.~5. For
$\widetilde{\chi} \widetilde{\chi} \rightarrow Z Z$,
$\widetilde{\chi} \widetilde{\chi} \rightarrow W^+ W^-$ and
$\widetilde{\chi} \widetilde{\chi} \rightarrow \tau^+ \tau^-$ channel,
cross sections are displayed in Fig.~6 and Fig.~7.  It is revealed
that these cross sections increase significantly due to Sommerfeld
Enhancement in these figures.
In Fig.~9, we compare our calculation result with the limits of 
current experimental data. Clearly, we can constrain the range of
prohibited dark matter mass with the value of $\delta m = 3~{\rm MeV}$.
We vary the value of $\delta m$ from $3~{\rm MeV}$ to $4~{\rm MeV}$
then of course we get a similar limit comparing with the current
experimental result as in Fig.~9. Further, we continue the same
processes to vary the value of $\delta m$ up to $10~{\rm MeV}$. Then
finally all constraints on $\delta m$ vs DM mass $m$ are plotted in Fig.~11.
On the left panel of Fig.~12, we show the limited parameter region 
acquired by comparing our result with future planned sensitivity of
CTA~\cite{Ong:2019zyq,Acharyya:2020sbj}. On the right panel of
Fig.~12, we draw the disallowed parameter band which is obtained by
comparison of the case of restricting about 100 times stronger
according to the future planned experiments \mbox{\it e.g.}
MAGIC~\cite{Ahnen:2017pqx}.

In conclusion the cross section calculated by our model has already
reached an observable range. Thus we can find signals from DM, and solve
the problems such as dark matter and Li problems in the near
future. In addition, even if no signal is detected, the parameters of
the model can be limited and the validity of the supersymmetric
particles can be verified soon.

\acknowledgments
This work was supported by JSPS KAKENHI Grants
No. JP18H01210 (J.S.), and MEXT KAKENHI Grant No. JP18H05543 (J.S.).

\newpage

\bibliographystyle{h-physrev5}
\bibliography{bunken}

\begin{thebibliography}{10}

\bibitem{Fukuda_1998}
Y.~Fukuda {\em et~al.},
\newblock Physical Review Letters {\bf 81}, 1562^^e2^^80^^931567 (1998).

\bibitem{Wittman:2000tc}
D.~M. Wittman, J.~Tyson, D.~Kirkman, I.~Dell'Antonio, and G.~Bernstein,
\newblock Nature {\bf 405}, 143 (2000), arXiv:astro-ph/0003014.

\bibitem{Komatsu:2010fb}
WMAP, E.~Komatsu {\em et~al.},
\newblock Astrophys. J. Suppl. {\bf 192}, 18 (2011), arXiv:1001.4538.

\bibitem{Aghanim:2018eyx}
Planck, N.~Aghanim {\em et~al.},
\newblock Astron. Astrophys. {\bf 641}, A6 (2020), arXiv:1807.06209.

\bibitem{Ellis:1983ew}
J.~R. Ellis, J.~Hagelin, D.~V. Nanopoulos, K.~A. Olive, and M.~Srednicki,
\newblock Nucl. Phys. B {\bf 238}, 453 (1984).

\bibitem{Jungman_1996}
G.~Jungman, M.~Kamionkowski, and K.~Griest,
\newblock Physics Reports {\bf 267}, 195^^e2^^80^^93373 (1996).

\bibitem{Ryan:2000zz}
S.~G. Ryan, T.~C. Beers, K.~A. Olive, B.~D. Fields, and J.~E. Norris,
\newblock Astrophys. J. Lett. {\bf 530}, L57 (2000), arXiv:astro-ph/9905211.

\bibitem{Sbordone:2010zi}
L.~Sbordone {\em et~al.},
\newblock Astron. Astrophys. {\bf 522}, A26 (2010), arXiv:1003.4510.

\bibitem{Cyburt:2008kw}
R.~H. Cyburt, B.~D. Fields, and K.~A. Olive,
\newblock JCAP {\bf 11}, 012 (2008), arXiv:0808.2818.

\bibitem{Jittoh:2007fr}
T.~Jittoh {\em et~al.},
\newblock Phys. Rev. D {\bf 76}, 125023 (2007), arXiv:0704.2914.

\bibitem{Jittoh:2008eq}
T.~Jittoh {\em et~al.},
\newblock Phys. Rev. D {\bf 78}, 055007 (2008), arXiv:0805.3389.

\bibitem{Jittoh:2010wh}
T.~Jittoh {\em et~al.},
\newblock Phys. Rev. D {\bf 82}, 115030 (2010), arXiv:1001.1217.

\bibitem{Jittoh:2011ni}
T.~Jittoh {\em et~al.},
\newblock Phys. Rev. D {\bf 84}, 035008 (2011), arXiv:1105.1431.

\bibitem{Kohri:2012gc}
K.~Kohri, S.~Ohta, J.~Sato, T.~Shimomura, and M.~Yamanaka,
\newblock Phys. Rev. D {\bf 86}, 095024 (2012), arXiv:1208.5533.

\bibitem{Konishi:2013gda}
Y.~Konishi {\em et~al.},
\newblock Phys. Rev. D {\bf 89}, 075006 (2014), arXiv:1309.2067.

\bibitem{Hisano_2007}
J.~Hisano, S.~Matsumot, M.~Nagai, O.~Saito, and M.~Senami,
\newblock Physics Letters B {\bf 646}, 34^^e2^^80^^9338 (2007).

\bibitem{Ellis:1999mm}
J.~R. Ellis, T.~Falk, K.~A. Olive, and M.~Srednicki,
\newblock Astropart. Phys. {\bf 13}, 181 (2000), arXiv:hep-ph/9905481,
\newblock [Erratum: Astropart.Phys. 15, 413--414 (2001)].

\bibitem{ArkaniHamed:2008qn}
N.~Arkani-Hamed, D.~P. Finkbeiner, T.~R. Slatyer, and N.~Weiner,
\newblock Phys. Rev. D {\bf 79}, 015014 (2009), arXiv:0810.0713.

\bibitem{Feng:2010zp}
J.~L. Feng, M.~Kaplinghat, and H.-B. Yu,
\newblock Phys. Rev. D {\bf 82}, 083525 (2010), arXiv:1005.4678.

\bibitem{Abramowski:2014tra}
H.E.S.S., A.~Abramowski {\em et~al.},
\newblock Phys. Rev. D {\bf 90}, 112012 (2014), arXiv:1410.2589.

\bibitem{Abdo:2009zk}
Fermi-LAT, A.~A. Abdo {\em et~al.},
\newblock Phys. Rev. Lett. {\bf 102}, 181101 (2009), arXiv:0905.0025.

\bibitem{Biondini:2018ovz}
S.~Biondini and S.~Vogl,
\newblock JHEP {\bf 02}, 016 (2019), arXiv:1811.02581.

\bibitem{ElAisati:2017ppn}
C.~El~Aisati, C.~Garcia-Cely, T.~Hambye, and L.~Vanderheyden,
\newblock JCAP {\bf 10}, 021 (2017), arXiv:1706.06600.

\bibitem{Abdalla:2018mve}
HESS, H.~Abdalla {\em et~al.},
\newblock JCAP {\bf 11}, 037 (2018), arXiv:1810.00995.

\bibitem{Braaten:2017kci}
E.~Braaten, E.~Johnson, and H.~Zhang,
\newblock JHEP {\bf 02}, 150 (2018), arXiv:1708.07155.

\bibitem{Lisanti:2016jxe}
M.~Lisanti,
\newblock {Lectures on Dark Matter Physics},
\newblock in {\em {Theoretical Advanced Study Institute in Elementary Particle
  Physics}: {New Frontiers in Fields and Strings}}, pp. 399--446, 2017,
  arXiv:1603.03797.

\bibitem{Hisano:2004ds}
J.~Hisano, S.~Matsumoto, M.~M. Nojiri, and O.~Saito,
\newblock Phys. Rev. D {\bf 71}, 063528 (2005), arXiv:hep-ph/0412403.

\bibitem{Matsumoto:2005ui}
S.~Matsumoto, J.~Sato, and Y.~Sato,
\newblock (2005), arXiv:hep-ph/0505160.

\bibitem{Bergstrom:1997fj}
L.~Bergstrom, P.~Ullio, and J.~H. Buckley,
\newblock Astropart. Phys. {\bf 9}, 137 (1998), arXiv:astro-ph/9712318.

\bibitem{Ackermann:2015zua}
Fermi-LAT, M.~Ackermann {\em et~al.},
\newblock Phys. Rev. Lett. {\bf 115}, 231301 (2015), arXiv:1503.02641.

\bibitem{Navarro:1996gj}
J.~F. Navarro, C.~S. Frenk, and S.~D.~M. White,
\newblock Astrophys. J. {\bf 490}, 493 (1997), arXiv:astro-ph/9611107.

\bibitem{Burkert:1995yz}
A.~Burkert,
\newblock IAU Symp. {\bf 171}, 175 (1996), arXiv:astro-ph/9504041.

\bibitem{Graham:2005xx}
A.~W. Graham, D.~Merritt, B.~Moore, J.~Diemand, and B.~Terzic,
\newblock Astron. J. {\bf 132}, 2685 (2006), arXiv:astro-ph/0509417.

\bibitem{Navarro:2008kc}
J.~F. Navarro {\em et~al.},
\newblock Mon. Not. Roy. Astron. Soc. {\bf 402}, 21 (2010), arXiv:0810.1522.

\bibitem{Cirelli:2010xx}
M.~Cirelli {\em et~al.},
\newblock JCAP {\bf 03}, 051 (2011), arXiv:1012.4515,
\newblock [Erratum: JCAP 10, E01 (2012)].

\bibitem{Sjostrand:2014zea}
T.~Sj\"ostrand {\em et~al.},
\newblock Comput. Phys. Commun. {\bf 191}, 159 (2015), arXiv:1410.3012.

\bibitem{Kubo:2018xrk}
M.~Kubo, J.~Sato, T.~Shimomura, Y.~Takanishi, and M.~Yamanaka,
\newblock Phys. Rev. D {\bf 97}, 115013 (2018), arXiv:1803.07686.

\bibitem{Aaboud:2019trc}
ATLAS, M.~Aaboud {\em et~al.},
\newblock Phys. Rev. D {\bf 99}, 092007 (2019), arXiv:1902.01636.

\bibitem{Abdallah:2018qtu}
HESS, H.~Abdallah {\em et~al.},
\newblock Phys. Rev. Lett. {\bf 120}, 201101 (2018), arXiv:1805.05741.

\bibitem{Hryczuk:2019nql}
A.~Hryczuk {\em et~al.},
\newblock JHEP {\bf 10}, 043 (2019), arXiv:1905.00315.

\bibitem{Ong:2019zyq}
CTA Consortium, R.~A. Ong,
\newblock EPJ Web Conf. {\bf 209}, 01038 (2019), arXiv:1904.12196.

\bibitem{Acharyya:2020sbj}
CTA, A.~Acharyya {\em et~al.},
\newblock (2020), arXiv:2007.16129.

\bibitem{Ahnen:2017pqx}
MAGIC, M.~L. Ahnen {\em et~al.},
\newblock JCAP {\bf 03}, 009 (2018), arXiv:1712.03095.

\end{thebibliography}

\newpage

\section*{Appendix A}
For Lagrangian in (\ref{lagrangian}), all terms are expressed as follows.
\begin{align}
\mathscr{L}_{KT} =&\frac{1}{2}\bar{\widetilde{\chi}}\left(i\Slash{\partial}-m\right)\widetilde{\chi} +\bar{\widetilde{C}}_\alpha \left(i\Slash{\partial}-m_{C_\alpha} \right)\widetilde{C}^\alpha+\bar{\nu}_D^i i\Slash{\partial}\nu_{Di} +\bar{e}^i \left(i\Slash{\partial}\delta_i^j-{(m_e)_i}^j\right){e}_j \nonumber \\
-&\widetilde{\tau}^*(\partial^2+m_{\widetilde{\tau}}^2)\widetilde{\tau} -\widetilde{\nu}^{*i}(\partial^2 +{{m_{\widetilde{\nu}}^2})\widetilde{\nu}_i +\frac{1}{2}Z_\mu (\partial^2+m_Z^2)}Z^\mu \nonumber \\
+&\frac{1}{2}A_\mu \partial^2A^\mu+W_\mu ^+(\partial^2+m_W^2) {W^\mu }^--\frac{1}{2}h^0 (\partial^2+m_{h^0}^2)h^0 ,
\end{align}
where $i,j=1,2,3$ and
\begin{align}
  \mathscr{L}_{\rm int}= \mathscr{L}_{\rm gauge}+\mathscr{L}_{h^0-\widetilde{\tau}-\widetilde{\tau}}+\mathscr{L}_{h^0-h^0-\widetilde{\tau}-\widetilde{\tau}}+\mathscr{L}_{\rm gaugino}
  + \mathscr{L}_{\rm chargino}.
\label{Lag}
\end{align}

The details of each interaction term except $\mathscr{L}_{gauge}$ are described below. In the following formula, the sums of $i$ are taken from 1 to 3 and we consider the lightest neutralino and slepton only. 

\begin{enumerate}
\item[$\bullet$] Higgs-stau-stau 3-point interaction
\begin{align}
& \mathscr{L}_{h^0-\widetilde{\tau}-\widetilde{\tau}} \nonumber \\
&= \frac{1}{\sqrt{2}}\widetilde{\tau}^* \widetilde{\tau} h^0
\left[ -\frac{1}{2}(c_\alpha v_u + s_\alpha v_d)
\left( g^2 N_{\widetilde{l}}{_1}^i N^\dagger_{\widetilde{l}}{_i}^1 +
g'{^2} \left( -N_{\widetilde{l}}{_1}^i N^\dagger_{\widetilde{l}}{_i}^1
+2 N_{\widetilde{l}}{_1}^{i+3} N^\dagger_{\widetilde{l}}{_{i+3}}^1\right) \right) \right. \nonumber \\
& \qquad \qquad \qquad \quad +2 s_\alpha v_d \left( N_{\widetilde{l}}{_1}^i y{^\dagger}{_i}^k y{_k}^j N^\dagger_{\widetilde{l}}{_j}^1  
+ N_{\widetilde{l}}{_1}^{i+3} y{_i}^k y{^\dagger}{_k}^j N^\dagger_{\widetilde{l}}{_{j+3}}^1 \right)  \nonumber \\
& \left. \qquad \qquad \qquad \quad +c_\alpha \left( \mu^* N_{\widetilde{l}}{_1}^{i+3} y{_i}^j N^\dagger_{\widetilde{l}}{_j}^1 + h.c.\right) + s_\alpha \left( N_{\widetilde{l}}{_1}^{i+3} a{_i}^j N^\dagger_{\widetilde{l}}{_j}^1 +h.c. \right) \right] .
\end{align}
  \item[$\bullet$] Higgs-stau-stau 4-point interaction
\begin{align}
&\mathscr{L}_{h^0-h^0-\widetilde{\tau}-\widetilde{\tau}} \nonumber \\
&=\widetilde{\tau}^* \widetilde{\tau} {h^0}^2 \left[-\frac{1}{8}g^2(c_\alpha^2-s_\alpha^2) N_{\widetilde{l}}{_1}^i N^\dagger_{\widetilde{l}}{_i}^1 
-\frac{1}{8}g'{^2}(c_\alpha^2-s_\alpha^2)\left( -N_{\widetilde{l}}{_1}^i N^\dagger_{\widetilde{l}}{_i}^1 +2N_{\widetilde{l}}{_1}^{i+3} N^\dagger_{\widetilde{l}}{_{i+3}}^1 \right) \right. \nonumber \\
&  \left. \qquad  \qquad \qquad -\frac{1}{2}s_\alpha^2 \left( N_{\widetilde{l}}{_1}^i y{^\dagger}{_i}^k y{_k}^j N^\dagger_{\widetilde{l}}{_j}^1  
+ N_{\widetilde{l}}{_1}^{i+3} y{_i}^k y{^\dagger}{_k}^j N^\dagger_{\widetilde{l}}{_{j+3}}^1 \right) \right] .
\end{align}
\item[$\bullet$] Gaugino-interaction
\begin{align}
\mathscr{L}_{\rm gaugino}=&\widetilde{\tau}^* \bar{\widetilde{\chi}} \left[ P_L \left(\frac{\sqrt{2}}{2}g' N_{\widetilde{l}}{_1}^{i} N_{\widetilde{G}}{_1}^1
+ \frac{\sqrt{2}}{2}g N_{\widetilde{l}}{_1}^{i} N_{\widetilde{G}}{_1}^2
+\frac{1}{2} N_{\widetilde{l}}{_1}^{j+3} {y_j}^i N_{\widetilde{G}}{_1}^4\right) \right. \nonumber \\
& \qquad \quad \left.  + P_R \left(-\sqrt{2}g' N_{\widetilde{l}}{_1}^{i+3} N_{\widetilde{G}}{_1}^1
+\frac{1}{2} N_{\widetilde{l}}{_1}^{i} y{^\dagger}{_i}^j N_{\widetilde{G}}{_1}^4\right) \right] {e_D}_i +h.c.
\end{align}
where, $P_L,P_R$ is projection operetor given as
\begin{align}
P_L=\frac{1-\gamma_5}{2}, \ \ \ \ P_R=\frac{1+\gamma_5}{2}.
\end{align}
Also, four-component spinors ${e_D}_i,\, \widetilde{\psi}^0_D{_a}$ are defined as
\begin{align*}
{e_D}_i=
\left(
\begin{array}{c}
{e_L}_\alpha{_i} \\
{e_R}^{\dot{\alpha}}_i \\
\end{array}
\right),\qquad
\widetilde{\psi}^0_D{_a}=
\left(
\begin{array}{c}
{\widetilde{\psi}^0}_\alpha{_a} \\
{{\widetilde{\psi}^0}{^\dagger}}{_a^{\dot{\alpha}}} \\
\end{array}
\right).
\end{align*}
\item[$\bullet$]Interaction between neutrino and chargino to release neutrino in the final state
\begin{align}
\mathscr{L}_{\rm chargino}=&-g \widetilde{e}_L^*{^i} \widetilde{W}^- \nu_i 
-g \nu^\dagger{^i} {\widetilde{W}^-}{^\dagger} \widetilde{e}_{Li}
-\frac{1}{2} \widetilde{e}_R^*{^i} {y_i}^j \nu_j \widetilde{H}^-_d
-\frac{1}{2} {\nu^{\dagger}}^i y{^\dagger}{_i}^j \widetilde{e}_R{_j} {\widetilde{H}^-_d}{^\dagger}  \nonumber \\
=&\widetilde{\tau}^* \bar{\widetilde{C}}_\alpha P_L \nu_{Di} 
\left[ -g N_{\widetilde{l}}{_1}^{i} {{U^\dagger}_1}^\alpha -\frac{1}{2} N_{\widetilde{l}}{_1}^{j+3} {y_j}^i {{U^\dagger}_2}^\alpha \right]
+h.c. ,
\end{align}
where four-component spinor $\widetilde{C}^\alpha$ is
\begin{gather*}
\widetilde{C}^\alpha=\left(
\begin{array}{c}
\widetilde{C}_\alpha^+ \\
{\widetilde{C}^-}{^{\dagger \alpha}} \\
\end{array}
\right) .
\end{gather*}
In Eq.(\ref{lagrangian}), main terms are chosen.

\end{enumerate}

\newpage

\section*{Appendix B}
We note the result of integrating out the fields other than $A_\mu$. In the following formula, the sums of $i$ is taken from 1 to 3.  \\
\begin{enumerate}
\item[$\bullet$] Effective action obtained by integrating out $Z_\mu$
\begin{align}
\S_Z =& 2ie^2 g_z^2 \left(s_w^2-\frac{1}{2} N_{\widetilde{l}}{_1}^i N^\dagger_{\widetilde{l}}{_i}^1\right)^2 \nonumber \\
& \qquad \times {\rm tr}\int d^4 x_1 d^4 x_2 |\widetilde{\tau}|^2(x_1) |\widetilde{\tau}|^2(x_2) {D_{\mu \nu}^Z}(x_1-x_2) {D^{A \nu \rho}}(x_2-x_1) \nonumber \\
+&ig_z^4 \left(s_w^2-\frac{1}{2} N_{\widetilde{l}}{_1}^i N^\dagger_{\widetilde{l}}{_i}^1\right)^4 \nonumber \\
&\qquad \times {\rm tr} \int d^4 x_1 d^4 x_2 |\widetilde{\tau}|^2(x_1) |\widetilde{\tau}|^2(x_2) {D_{\mu \nu}^Z}(x_1-x_2) D^{Z \nu \rho }(x_2-x_1) \nonumber \\
+& \frac{i}{2}\int d^4 x d^4 y J^\mu_Z(x) {D_{\mu \nu}^Z}(x-y) J^\nu_Z(y),
\end{align}
where
\begin{align}
{D_{\mu \nu}^Z}(x-y)=&-i\int \frac{d^4q}{{(2\pi)}^4}\frac{g_{\mu \nu}}{q^2-m_Z^2+i\epsilon}e^{-iq(x-y)} ,\\
J^\mu_Z(x)=&-i g_z  \left(s_w^2-\frac{1}{2} N_{\widetilde{l}}{_1}^i N^\dagger_{\widetilde{l}}{_i}^1\right) \widetilde{\tau}^* \overleftrightarrow{\partial}^\mu  \widetilde{\tau}.
\end{align}
\item[$\bullet$]Effective action obtained by integrating out $W^+,W^-$
\begin{align}
\S_W=&\frac{i}{4} g^4 \left( N_{\widetilde{l}}{_1}^i N^\dagger_{\widetilde{l}}{_i}^1 \right) ^2 {\rm tr} \int d^4 x_1 d^4 x_2 |\widetilde{\tau}|^2(x_1) |\widetilde{\tau}|^2(x_2) {D_{\mu \nu}^W}(x_1-x_2) D^{W \nu \rho }(x_2-x_1) \nonumber \\
&+i\int d^4 x d^4 y J^\mu_W(x) {D_{\mu \nu}^W}(x-y) {J^\nu_W}^\dagger (y),
\end{align}
where
\begin{align}
{D_{\mu \nu}^W}(x-y)=&-i\int \frac{d^4q}{{(2\pi)}^4}\frac{g_{\mu \nu}}{q^2-m_W^2+i\epsilon}e^{-iq(x-y)} ,\\
J^\mu_W(x)=& -i\frac{\sqrt{2}}{2}g N_{\widetilde{l}}{_1}^i \widetilde{\tau}^*  \overleftrightarrow{\partial}^\mu  \widetilde{\nu}{_i} , \\
{J^\mu_W}^\dagger (x) =&-i\frac{\sqrt{2}}{2}g N^\dagger_{\widetilde{l}}{_i}^1 \widetilde{\nu}{^*}{^i} \overleftrightarrow{\partial}^\mu \widetilde{\tau} .
\end{align}
\item[$\bullet$]Effective action obtained by integrating out $\widetilde{\nu}{^*}{^i},\widetilde{\nu}{_i}$
\begin{align}
\S_{\widetilde{\nu}}=&2ig^4 N^\dagger_{\widetilde{l}}{_i}^1 N_{\widetilde{l}}{_1}^j N^\dagger_{\widetilde{l}}{_k}^1 N_{\widetilde{l}}{_1}^l {\rm tr}\int d^4 x_1 d^4 x_2 d^4 x_3 d^4 x_4 \nonumber \\
& \qquad \times \partial ^\sigma \widetilde{\tau}^*(x_1) \partial ^\mu \widetilde{\tau}(x_2) \partial ^\nu \widetilde{\tau}^*(x_3) \partial ^\rho \widetilde{\tau}(x_4) \nonumber \\
& \qquad \times{D^{\widetilde{\nu}}}_j{^i} (x_1-x_2) {D_{\mu \nu}^W}(x_3-x_2) {D^{\widetilde{\nu}}}_l{^k}(x_3-x_4) {D_{\rho \sigma}^W}(x_1-x_4),
\end{align}
where
\begin{align}
{D^{\widetilde{\nu}}}_i{^j}(x-y)=-i\int \frac{d^4q}{{(2\pi)}^4}\frac{\delta _i^j}{q^2-m_{\widetilde{\nu}}^2+i\epsilon}e^{-iq(x-y)}.
\end{align}
\item[$\bullet$]Effective action obtained by integrating out $e_D,{\bar{e}}_D$
\begin{align}
{\cal S}_e=i\int d^4 x d^4 y \widetilde{\tau}^*(x) \widetilde{\tau}(y) {\bar{\widetilde{\chi}}}(x)\left[ C^i_1 P_L +C^i_2 P_R \right] {S^{\tau}(x-y)}_i{^j} \left[ {C^\dagger}_{1j} P_R +{C^\dagger}_{2j} P_L \right]{\widetilde{\chi}}(y),
\end{align}
where
\begin{align}
{S^{\tau}(x-y)}_i{^j}=&i\int \frac{d^4q}{{(2\pi)}^4}\frac{\Slash{q}\delta_i^j+{(m_e)_i}^j}{q^2-m_e^2+i\epsilon}e^{-iq(x-y)} ,\\
C^i_1=&\left(\frac{\sqrt{2}}{2}g' N_{\widetilde{l}}{_1}^{i} N_{\widetilde{G}}{_1}^1
+ \frac{\sqrt{2}}{2}g N_{\widetilde{l}}{_1}^{i} N_{\widetilde{G}}{_1}^2
+\frac{1}{2} N_{\widetilde{l}}{_1}^{j+3} {y_j}^i N_{\widetilde{G}}{_1}^4\right) , \label{C1}\\
C^i_2=&\left(-\sqrt{2}g' N_{\widetilde{l}}{_1}^{i+3} N_{\widetilde{G}}{_1}^1
+\frac{1}{2} N_{\widetilde{l}}{_1}^{i} y{^\dagger}{_i}^1 N_{\widetilde{G}}{_1}^4\right).\label{C2}
\end{align}
\item[$\bullet$]Effective action obtained by integrating out $\nu_D,\bar{\nu}_D$
\begin{align}
{\cal S}_{\widetilde{\nu}} =i\int d^4 x d^4 y \widetilde{\tau}^*(x) \widetilde{\tau}(y) \bar{\widetilde{C}}_\alpha (x) C^{i \alpha} P_L {S^{\nu}(x-y)}_i{^j} C^\dagger_{j \beta} P_R {\widetilde{C}}^\beta(y),
\end{align}
where
\begin{align}
{S^{\nu}(x-y)}_i{^j}=i\int \frac{d^4q}{{(2\pi)}^4}\frac{\Slash{q}\delta_i^j}{q^2+i\epsilon}e^{-iq(x-y)} , \\
C^{i \alpha}=-g N_{\widetilde{l}}{_1}^{i} {{U^\dagger}_1}^\alpha -\frac{1}{2} N_{\widetilde{l}}{_1}^{j+3} {y_j}^i {{U^\dagger}_2}^\alpha.
\end{align}
\

\item[$\bullet$]Effective action obtained by integrating out $\bar{\widetilde{C}},\widetilde{C}$
\begin{align}
\S_{\widetilde{C}}&=\frac{i}{2}{\rm tr} \int d^4 x_1 d^4 x_2 d^4 x_3 d^4 x_4 \widetilde{\tau}^*(x_1) \widetilde{\tau}(x_2) \widetilde{\tau}^*(x_3) \widetilde{\tau}(x_4) \nonumber \\
\times& S^{\widetilde{C}}(x_1-x_2)_\alpha {^\beta} C^{i \alpha} P_L {S^{\nu}(x_2-x_3)}_i{^j} C^\dagger_{j \beta} P_R  S^{\widetilde{C}}(x_3-x_4)_\gamma {^\delta} C^{k \gamma} P_L {S^{\nu}(x_4-x_1)}_k{^l} C^\dagger_{l \delta} P_R,
\end{align}
where
\begin{align}
{S^{\widetilde{C}}(x-y)}_\alpha{^\beta}=i\int \frac{d^4q}{{(2\pi)}^4}\frac{\Slash{q}+m_{C_\alpha}}{q^2-m_{C_\alpha}^2+i\epsilon} \delta _\alpha^\beta e^{-iq(x-y)}.
\end{align}

\item[$\bullet$]Effective action obtained by integrating out $h^0$
\begin{align}
{\cal S}_{h^0}=&i {C_{h^0}^{(4)}}^2 {\rm tr} \int d^4 x_1 d^4 x_2 |\widetilde{\tau}|^2(x_1) |\widetilde{\tau}|^2(x_2) D^{h^0}(x_1-x_2) D^{h^0}(x_2-x_1) \nonumber \\
&-\frac{i}{2}\int d^4 x d^4 y J^{h^0}(x) D^{h^0}(x-y) J^{h^0}(y),
\end{align}
where
\begin{gather*}
D^{h^0}(x-y)=-i\int \frac{d^4q}{{(2\pi)}^4}\frac{1}{q^2-m_{h^0}^2+i\epsilon}e^{-iq(x-y)} , \\
J^{h^0}(x)=|{\widetilde{\tau}}(x)|^2 C_{h^0},
\end{gather*}
\begin{align}
C_{h^0}^{(4)}=&-\frac{1}{8}g^2(c_\alpha^2-s_\alpha^2) N_{\widetilde{l}}{_1}^i N^\dagger_{\widetilde{l}}{_i}^1 
-\frac{1}{8}g'{^2}(c_\alpha^2-s_\alpha^2)\left( -N_{\widetilde{l}}{_1}^i N^\dagger_{\widetilde{l}}{_i}^1 +2N_{\widetilde{l}}{_1}^{i+3} N^\dagger_{\widetilde{l}}{_{i+3}}^1 \right)  \nonumber \\
&\qquad \qquad  -\frac{1}{2}s_\alpha^2 \left( N_{\widetilde{l}}{_1}^i y{^\dagger}{_i}^k y{_k}^j N^\dagger_{\widetilde{l}}{_j}^1  
+ N_{\widetilde{l}}{_1}^{i+3} y{_i}^k y{^\dagger}{_k}^j N^\dagger_{\widetilde{l}}{_{j+3}}^1 \right), \\
C_{h^0}=&\frac{1}{\sqrt{2}}\left[ -\frac{1}{2}(c_\alpha v_u + s_\alpha v_d)
\left( g^2 N_{\widetilde{l}}{_1}^i N^\dagger_{\widetilde{l}}{_i}^1 +
g'{^2} \left( -N_{\widetilde{l}}{_1}^i N^\dagger_{\widetilde{l}}{_i}^1
+2 N_{\widetilde{l}}{_1}^{i+3} N^\dagger_{\widetilde{l}}{_{i+3}}^1\right) \right) \right. \nonumber \\
& \qquad \qquad \ \ +2 s_\alpha v_d \left( N_{\widetilde{l}}{_1}^i y{^\dagger}{_i}^k y{_k}^j N^\dagger_{\widetilde{l}}{_j}^1  
+ N_{\widetilde{l}}{_1}^{i+3} y{_i}^k y{^\dagger}{_k}^j N^\dagger_{\widetilde{l}}{_{j+3}}^1 \right)  \nonumber \\
& \qquad \qquad \ \ + \left.c_\alpha \left( \mu^* N_{\widetilde{l}}{_1}^{i+3} y{_i}^j N^\dagger_{\widetilde{l}}{_j}^1 + h.c.\right)  s_\alpha \left( N_{\widetilde{l}}{_1}^{i+3} a{_i}^j N^\dagger_{\widetilde{l}}{_j}^1 +h.c. \right) \right] \label{Ch0}. 
\end{align}

\end{enumerate}

Furthermore, we note 
\begin{align}
\S_A'=& i e^4 {\rm tr} \int d^4 x_1 d^4 x_2 |\widetilde{\tau}|^2(x_1) |\widetilde{\tau}|^2(x_2) {D_{\mu \nu}^A}(x_1-x_2)D^{A \nu \rho }(x_2-x_1) \nonumber \\
&+ \frac{i}{2}\int d^4 x d^4 y  J_A^\mu(x) {D_{\mu \nu}^A}(x-y)  J_A^\nu(y), \\
\S_W'=&\frac{i}{4} g^4 \left( N_{\widetilde{l}}{_1}^i N^\dagger_{\widetilde{l}}{_i}^1 \right) ^2 {\rm tr} \int d^4 x_1 d^4 x_2 |\widetilde{\tau}|^2(x_1) |\widetilde{\tau}|^2(x_2) {D_{\mu \nu}^W}(x_1-x_2) D^{W \nu \rho }(x_2-x_1),
\end{align}
where
\begin{align}
J_A^\mu(x) = ie\widetilde{\tau}^* \overleftrightarrow{\partial}^\mu  \widetilde{\tau}.
\end{align}

\newpage

\section*{Appendix C}
For the calculation of the imaginary part, the result is as follows for the fields other than $\S_{\gamma}$. For example,  a diagram in which fermions mediate is shown the below diagram.\\
\begin{fmffile}{boxtau}
\begin{center}
\parbox{30mm}{\begin{fmfgraph*}(40,30)
\fmftop{i1,o1}
\fmfbottom{i2,o2}
\fmf{dashes_arrow,label=$\widetilde{\tau}$}{i1,v1}
\fmf{fermion,label=$f$}{v1,v3}
\fmf{dashes_arrow,label=$\widetilde{\tau}$}{v3,o1}
\fmf{dashes_arrow,label=$\widetilde{\tau}$}{i2,v2}
\fmf{fermion,label=$f$}{v2,v4}
\fmf{dashes_arrow,label=$\widetilde{\tau}$}{v4,o2}
\fmf{fermion,label=$\widetilde{\chi}$,tension=0}{v1,v2}
\fmf{fermion,label=$\widetilde{\chi}$,tension=0}{v4,v3}
\end{fmfgraph*}}
\end{center}
\end{fmffile}

\ 

\

With optical theorem, effective action of this case becomes
\begin{align}
\S_{e}=&~ \frac{i}{16\pi} \frac{\left(1-m_e^2 / m^2 \right)^{3/2}}{(2m^2 - m_e^2)^2} \left[ (|C_1|^2+|C_2|^2 )(m_e^2+m(m_c+m_c^\dagger))+m_c m_c^\dagger + m^2 |C_1|^2 |C_2|^2)\right] \nonumber \\
&\times \int d^4 x \eta^*(x) \eta(x)\xi^*(x)\xi(x) \nonumber \\
&\equiv ~i~\Gamma_{e_i e_i}  \int d^4 x \eta^*(x) \eta(x)\xi^*(x)\xi(x) .
\end{align}
For other fields,
\begin{align}
\S_{Z}=& ~i~\left[ \frac{2 g_z^4\left(s_w^2-\frac{1}{2} {N_{\widetilde{l}1}}^i {{N^\dagger}_{\widetilde{l}i}}^1\right)^4 m^6}{\pi m_z^4} \frac{\left(1-\frac{m^2}{m^2}\right)^{\frac{5}{2}}}{\left(m^2+m_{\widetilde{\tau}}^2-m_z^2\right)^2} + \frac{g_z^4 (s_w^2 - \frac{1}{2} N_{\widetilde{l}1}^i {{N^\dagger}_{\widetilde{l}i}}^1)^4}{8\pi m^2}\sqrt{1-\frac{m_z^2}{m^2}}   \right] \nonumber \\
& \times \int d^4 x \eta^*(x) \eta(x)\xi^*(x)\xi(x) \nonumber \\
 \equiv & ~i~\Gamma_{Z^0 Z^0} \int d^4 x \eta^*(x) \eta(x) \xi^*(x) \xi(x),
\end{align}
where the sums of $i$ is taken from 1 to 3.
\begin{align}
\S_{AZ}=&~i ~\frac{e^2 g_z^2 \left(s_w^2-\frac{1}{2} N_{\widetilde{l}}{_1}^i N^\dagger_{\widetilde{l}}{_i}^1\right)^2} {4 \pi m^2} \left( 1-\frac{m_z^2}{4m^2}\right) \int d^4 x \eta^*(x) \eta(x) \xi^*(x) \xi(x)\nonumber \\
\equiv & ~ \Gamma_{\gamma Z^0} \int d^4 x \eta^*(x) \eta(x) \xi^*(x) \xi(x) .
\end{align}

\begin{align}
\S_{W}=~&i~\left[ \frac{g^4\left({N_{\widetilde{l}1}}^i {{N^\dagger}_{\widetilde{l}i}}^1\right)^4 m^6}{8 \pi m_W^4} \left(1-\frac{m_W^2}{m^2}\right)^\frac{5}{2} \frac{1}{\left(m^2+m_{\widetilde{\nu}}^2-m_W^2\right)^2}+\frac{g^4 \left(N_{\widetilde{l}1}^i {{N^\dagger}_{\widetilde{l}i}}^1\right)^4}{16\pi m^2}\sqrt{1-\frac{m_W^2}{m^2}}\right] \nonumber \\
&\times \int d^4 x \eta^*(x) \eta(x)\xi^*(x)\xi(x) \nonumber \\
\equiv& ~i~\Gamma_{W^+W^-}  \int d^4 x \eta^*(x) \eta(x)\xi^*(x)\xi(x) .
\end{align}

\begin{align}
\S_{\nu}=~i~&\frac{\left(C^{i\alpha} C_{i\alpha}^\dagger\right)^2}{32\pi} \frac{m_{C_\alpha}^2}{\left(m^2+m_{C_\alpha}^2 \right)^2}\nonumber \int d^4 x \eta^*(x) \eta(x)\xi^*(x)\xi(x) \nonumber \\
\equiv& ~i~\Gamma_{\nu_i \nu_i}  \int d^4 x \eta^*(x) \eta(x)\xi^*(x)\xi(x) .
\end{align}

\begin{align}
\S_{h^0} &=~ i\left[\frac{{C_{h^0}}^4}{4\pi m^2}\sqrt{1-\frac{m_{h^0}^2}{m^2}} \frac{1}{(m^2+m_{\widetilde{\tau}}^2 - m_{h^0}^2)^2}+\frac{{C_{h^0}^{(4)}}^4}{32 \pi m^2} \sqrt{1-\frac{m_{h^0}^2}{m^2}}\right] \nonumber \\
&\qquad \qquad \times \int d^4 x \eta^*(x) \eta(x)\xi^*(x)\xi(x) \nonumber \\
\equiv&~ i~\Gamma_{h^0 h^0}  \int d^4 x \eta^*(x) \eta(x)\xi^*(x)\xi(x) .
\end{align}

\end{document}